\documentclass[12pt,a4paper]{article}
\usepackage[a4paper]{geometry}

\usepackage{amsmath}
\usepackage{amsfonts}
\usepackage{amssymb}
\usepackage{dsfont}
\usepackage{graphicx}
\usepackage[nosort]{cite}
\usepackage{stmaryrd}
\usepackage{mathtools}

\allowdisplaybreaks[2]
\numberwithin{equation}{section}

\newcommand{\eq}[1]{\begin{equation}
                     \begin{split} #1 \end{split}
                     \end{equation}}

\newcommand{\ov}[1]{\overline{#1}}
\newcommand{\ul}[1]{\underline{#1}}
\newcommand{\op}{\hspace{1pt}}

\newcommand{\scalemath}[2]{\scalebox{#1}{\mbox{\ensuremath{\displaystyle #2}}}}

\begin{document}


%

\vspace*{1.5cm}


\begin{center}
{\LARGE
On T-duality transformations \\[5pt] for the three-sphere
}
\end{center}


\vspace{0.4cm}

\begin{center}
  Erik Plauschinn
\end{center}


\vspace{0.4cm}

\begin{center} 
\emph{Dipartimento di Fisica e Astronomia ``Galileo Galilei'' \\
Universit\`a  di Padova \\ Via Marzolo 8, 35131 Padova, Italy}  \\[0.1cm] 
\vspace{0.4cm}
and \\[0.1cm]
\vspace{0.4cm}
\emph{INFN, Sezione di Padova \\
Via Marzolo 8, 35131 Padova, Italy}  \\
\end{center} 

\vspace{2.1cm}


\begin{abstract}
\noindent
We study collective T-duality transformations along one, two and three 
directions of isometry for the three-sphere with $H$-flux. Our aim is to obtain new 
non-geometric backgrounds along lines similar to the example of the three-torus.
However, the resulting backgrounds turn out to be geometric in nature.
To perform the duality transformations, we develop a novel procedure for non-abelian
T-duality, which follows a route different compared to the known literature, and
which highlights the underlying structure from an alternative point of view.
\end{abstract}

\clearpage


\tableofcontents


\section{Introduction}

String theory is a theory of extended objects, which distinguishes it from  ordinary quantum field 
theories of point particles. In particular, string theory contains closed strings, for which two 
types of excitations can be found in the spectrum: left-moving and right-moving modes.
When a closed string is probing a background
in which these two sectors behave in the same way, roughly speaking, both sectors ``see'' the same 
geometry. Hence,  one can give a geometric interpretation
of the background (at least in the large volume regime). 
However, in general the left- and right-moving sectors do not
need to be the same, but can detect the background differently. In this case, 
no geometric description is available and the corresponding background is called 
non-geometric.

Usually, string theory is studied in the geometric regime for which 
a large variety of background spaces is known, however,  in the non-geometric setting
it is more difficult to obtain explicit examples. 
One of the strategies to construct backgrounds for the non-geometric case is to apply T-duality transformations
to a known geometric space with non-vanishing NS-NS field strength $H$.
The prime example for this approach \cite{Shelton:2005cf} is the 
flat three-torus with $H\neq0$, leading to
\eq{
  \label{t_chain}
  H_{xyz} \quad\xleftrightarrow{\;\;\; T_{z}\;\;\;}\quad
   f_{xy}{}^{z} \quad\xleftrightarrow{\;\;\; T_{y}\;\;\;}\quad
  Q_{x}{}^{yz} \quad\xleftrightarrow{\;\;\; T_{x}\;\;\;}\quad
  R^{xyz} \; ,
}
where this  chain of T-duality transformations 
can be explained as follows. 
\begin{itemize}

\item The starting point is a flat three-torus with non-trivial $H$-flux, 
on which one performs a first T-duality transformation.
This results in a twisted torus
with vanishing field strength,  where the topology is characterized  by a so-called 
geometric flux $f$ \cite{Dasgupta:1999ss,Kachru:2002sk}.

\item A second  T-duality transformation  leads to a background with a locally-geometric description, 
which is however globally non-geometric \cite{Hellerman:2002ax}. 
The latter means that when considering a covering of the torus by open neighborhoods, the 
transition functions on the overlap of the charts are not solely given by diffeomorphisms,
and hence such a manifold cannot be described by Riemannian geometry.
However,  if in addition to diffeomorphisms 
one includes T-duality transformations as transition maps \cite{Dabholkar:2002sy},
this space can be globally defined.
This construction is called a T-fold \cite{Hull:2004in}, and carries a  so-called $Q$-flux \cite{Shelton:2005cf}. 
The $Q$-flux is related to non-commutative features of this background, and non-commutativity
in this context has been studied 
for instance in
\cite{Mathai:2004qq,Mathai:2004qc,Grange:2006es,Lust:2010iy,Lust:2012fp,Condeescu:2012sp,Chatzistavrakidis:2012qj,Andriot:2012vb,Bakas:2013jwa,Blair:2014kla},
and has been reviewed recently in \cite{Blumenhagen:2014sba}.

\item It has also been argued that formally a third T-duality transformation can be performed \cite{Shelton:2005cf},
but the resulting $R$-flux background is not even locally geometric and exhibits a non-associative structure.
These spaces have been studied from a  mathematical point of view in 
\cite{Bouwknegt:2004ap,Bouwknegt:2004tr}, later in \cite{Ellwood:2006my}, and have
been reconsidered in a series of papers \cite{Blumenhagen:2010hj,Blumenhagen:2011ph,Blumenhagen:2011yv,Lust:2012fp,Mylonas:2012pg,Plauschinn:2012kd,Chatzistavrakidis:2012qj,Bakas:2013jwa,
Deser:2013pra,Mylonas:2013jha,Blair:2014kla} more recently. 
A review from a mathematical perspective can be found in
\cite{Mylonas:2014aga}.

\end{itemize}
Another class of backgrounds showing non-geometric features are asymmetric 
orbifolds. In the context of non-geometry these have been studied for instance in 
 \cite{Dabholkar:2002sy,Hellerman:2002ax,Flournoy:2004vn,Flournoy:2005xe,Hellerman:2006tx,Blumenhagen:2011ph,Condeescu:2012sp,Condeescu:2013yma},
but they will not be the focus of this work.

There are a number of different approaches to investigate non-geometric backgrounds.
In addition to the above-mentioned line of research, we note that non-geometric flux configurations 
have been studied from a doubled-geometry point of view in \cite{Hull:2004in,Dabholkar:2005ve,Hull:2006va}.
More recently, non-geometric backgrounds have been investigated via field redefinitions
for the ten-dimensional supergravity action in 
\cite{Andriot:2011uh,Andriot:2012wx,Andriot:2012an,Blumenhagen:2012nk,Blumenhagen:2012nt,Blumenhagen:2013aia,Andriot:2013xca,Andriot:2014uda}, and 
have been analyzed from a world-sheet point of view for instance in 
\cite{Flournoy:2005xe,Halmagyi:2008dr,Halmagyi:2009te,Rennecke:2014sca}.
Also, there exists an extensive literature for non-geometry in the context of  double-field theory,
for which we would like to refer the reader to the reviews \cite{Aldazabal:2013sca,Hohm:2013bwa}.

\bigskip
The main purpose of the present paper is to study the chain of T-duality transformations
shown in \eqref{t_chain} not for the three-torus, but for the three-sphere with $H$-flux. 
One of the appealing features of the latter is that, in contrast to the torus, the string equations of 
motion can be  solved when the flux is appropriately adjusted. The main question we
want to answer is the following:
\begin{quote}
When applying two T-duality transformations to the three-sphere with $H$-flux, 
does one obtain a non-geometric $Q$-flux background?
\end{quote}
In order to address this point, a proper understanding of 
T-duality transformations is required. More concretely, since the isometry group of the three-sphere is non-abelian, 
we would like to be able to perform non-abelian T-duality transformations. 
These have been studied extensively in the past and some of the corresponding references are
\cite{delaOssa:1992vc,Giveon:1993ai,Alvarez:1993qi,Sfetsos:1994vz,Alvarez:1994np,Klimcik:1995ux,Lozano:1995jx,Curtright:1996ig};
more recently non-abelian T-duality has been discussed for instance in 
\cite{Sfetsos:2010uq,Itsios:2012dc,Itsios:2013wd,Sfetsos:2013wia}.
However, 
in this paper we are going to approach non-abelian T-duality from a slightly different point of view, which highlights
some of the structure important for our purposes. 
Let us furthermore mention that  some of the examples we will be discussing are related to results known 
in the literature;
nevertheless, our investigation here is in view of the chain of T-duality transformations shown in
equation \eqref{t_chain}.

\bigskip
This paper is organized as follows: in sections \ref{sec_prelim} and \ref{sec_nonab_t}
we develop a novel formalism for studying collective, and more generally non-abelian, T-duality transformations.
Our approach is based on \cite{Plauschinn:2013wta}, which for instance does not
require a gauge-fixing procedure and which is not based on Wess-Zumino-Witten models.
Furthermore, we are able to make explicit a particular constraint, shown in equation \eqref{variations_45},
which explains some of the structure found in the context of non-geometric backgrounds.

In section \ref{ex_1_torus} we apply collective T-duality transformations to the 
well-known example of the three-torus, thereby illustrating and checking our formalism.
In section \ref{sec_sphere} we study the chain of T-dualities \eqref{t_chain}
for the example of the three-sphere with $H$-flux; we find that 
after two T-duality transformations not a non-geometric but a geometric background is obtained. 

In section \ref{sec_summary} we summarize and discuss our findings, and in appendix
\ref{app} we collect results on collective (and non-abelian) T-duality transformations for the twisted three-torus
with $H$-flux.


\clearpage
\section{Preliminaries: non-linear sigma-model}
\label{sec_prelim}

We beginn our discussion by reviewing the sigma-model action 
for the NS-NS sector of the closed string, which 
encodes the dynamics of a target-space metric $G$, an anti-symmetric 
Kalb-Ramond field $B$, and a dilaton $\phi$.
In the second part of this section, we study gaugings of this action, thereby generalizing some results of  
\cite{Hull:1989jk,Hull:1990ms,Alvarez:1993qi,Hull:2006qs}


\subsubsection*{The action}

The sigma model is usually defined on a compact two-di\-men\-sion\-al manifold without boundaries,
corresponding to the world-sheet of a closed string.
However, in order to incorporate non-trivial field strengths $H=dB\neq0$ 
for the Kalb-Ramond field $B$, it turns out to be convenient to work with 
a Wess-Zumino term, which is defined on a compact  three-dimensional Euclidean world-sheet $\Sigma$ 
with two-dimensional boundary $\partial \Sigma$.
In this case, the sigma-model action takes the form
\eq{
  \label{action_01}
  \mathcal S =& -\frac{1}{4\pi \alpha'} \int_{\partial\Sigma} 
  \Bigl[ G_{ij} \, d X^i\wedge\star d X^j
  + \alpha'R\, \phi \star 1 \Bigr] \\[2mm]
  &-\frac{i}{2\pi \alpha'} \int_{\Sigma} \tfrac{1}{3!}\, H_{ijk}\op dX^i\wedge dX^j\wedge dX^k
  \,,
}
where the Hodge-star operator $\star$ is defined on  $\partial\Sigma$,  
and the differential is understood as $dX^i(\sigma^{\alpha}) = \partial_{\alpha} X^i d\sigma^{\alpha}$ 
with $\{\sigma^{\alpha}\}$ coordinates on $\partial \Sigma$ and on $\Sigma$.
The indices take values $i,j\in\{ 1,\ldots, d\}$ with $d$ the dimension of the target space,
and  $R$ denotes the curvature scalar corresponding to the world-sheet metric $h_{\alpha\beta}$
on $\partial\Sigma$.

Note that the choice of three-manifold $\Sigma$ for a given boundary $\partial\Sigma$ is
not unique. However, if the field strength $H$ is quantized, the path integral only depends on
the data of the two-dimensional theory \cite{Witten:1983tw}. In the above conventions, the quantization 
condition reads
\eq{
  \label{quantization}
  \frac{1}{2\pi\alpha'} \int_{\Sigma} H \;\in\; 2\pi \op\mathbb Z \,.
}


\subsubsection*{Symmetries of the world-sheet action}

The classical world-sheet action \eqref{action_01} is invariant under the standard world-sheet 
diffeomorphisms, but it can also have pure target-space symmetries of the form
\eq{
  \label{iso_trafo_01}
  \delta_{\epsilon} X^i = \epsilon^{\alpha}\op k_{\alpha}^i(X) 
}
for $\epsilon^{\alpha}$ constant, provided that three requirements are satisfied.
First, $k_{\alpha}$ with $\alpha=1,\ldots, N$ are  Killing vectors of the metric $G=G_{ij}\op dX^i\wedge \star dX^j$.
Second, there exist  one-forms $v_{\alpha}$ such that $\iota_{k_{\alpha}} H = dv_{\alpha}$ 
\cite{Hull:1989jk,Hull:1990ms}, 
and third, the Lie derivative of the dilaton $\phi$ in the direction of $k_{\alpha}$ vanishes. 
In terms of equations, these three conditions can be summarized as
\eq{
  \label{constraints_35}
  \mathcal L_{k_{\alpha}}  G = 0\,, \hspace{70pt}
  \iota_{k_{\alpha}} H = dv_{\alpha} \,, \hspace{70pt}
  \mathcal L_{k_{\alpha}} \phi =0\,,
}
where the Lie derivative is given by $\mathcal L_k = d\circ\iota_k + \iota_k\circ d$.
We also note that the  isometry algebra generated by the Killing vectors  is in general non-abelian
with structure constants $f_{\alpha\beta}{}^{\gamma}$,
\eq{
   \label{constraints_351}
   \bigl[ k_{\alpha} , k_{\beta} \bigr]_{\rm L} = f_{\alpha\beta}{}^{\gamma} \op k_{\gamma} \,.
}


\subsubsection*{Gauging a symmetry}

Let us now promote the global symmetries \eqref{iso_trafo_01} to  local
ones, with $\epsilon^{\alpha}$ depending on the world-sheet coordinates $\{\sigma^{\alpha}\}$. 
To do so, we introduce world-sheet gauge fields $A^{\alpha}$ and replace
$dX^i\to dX^i + k_{\alpha}^i A^{\alpha}$ for the term involving the metric. For the Wess-Zumino term $dX^i$ is 
kept unchanged, but 
additional  scalar fields $\chi_{\alpha}$ have to be introduced. The resulting gauged action  reads
\eq{
  \label{action_02}
  \widehat{\mathcal S} =&-\frac{1}{2\pi\alpha'} \int_{\partial\Sigma} \;  
  \tfrac{1}{2}\op G_{ij}  (dX^i + k^i_{\alpha} A^{\alpha})\wedge\star(dX^j + k^j_{\beta} A^{\beta})  
  \\[1mm]
  &-\frac{i}{2\pi \alpha'} \int_{\Sigma} \hspace{8pt} \tfrac{1}{3!}\, H_{ijk}\op dX^i\wedge dX^j\wedge dX^k
  \\[1mm]
  &-\frac{i}{2\pi \alpha'} \int_{\partial\Sigma} \:\Bigl[ \;
  ( v_{\alpha} + d\chi_{\alpha})\wedge A^{\alpha}
  + \tfrac{1}{2}\op \bigl( \iota_{k_{[\ul \alpha}} v_{\ul \beta]}
  + f_{\alpha\beta}{}^{\gamma} \chi_{\gamma} \bigr)\op A^{\alpha}\wedge A^{\beta}\;
  \Bigr] \,,
}
where we omitted the dilaton term, which does not get modified.
Now, given this action, there are two slightly different ways to implement the local symmetry 
transformations:
\begin{enumerate}

\item In the first approach, developed in detail in the two papers \cite{Hull:1989jk,Hull:1990ms}, 
the scalar fields $\chi_{\alpha}$ do not play a role; in fact, they are not mentioned at all.
In the present context, the local symmetry transformations then read as follows
\eq{
  \label{variantions_02}
  \hat\delta_{\epsilon} X^i = \epsilon^{\alpha} \op k^i_{\alpha} \,,\hspace{50pt}
  \hat \delta_{\epsilon} A^{\alpha} = - d\epsilon^{\alpha} - \epsilon^{\beta} A^{\gamma} f_{\beta\gamma}{}^{\alpha}
  \,,\hspace{50pt}
  \hat\delta_{\epsilon} \chi = 0 \,,
}
which have to be supplemented by the constraints\footnote{Our convention is that
the symmetrization and anti-symmetrization contains a factor of $1/n!$.}
\eq{
  \label{variantions_33}
  \mathcal L_{k_{[\ul \alpha}} v_{\ul \beta]} = f_{\alpha\beta}{}^{\gamma} v_{\gamma} \,,
  \hspace{70pt}
  \iota_{k_{(\ov \alpha}} v_{\ov \beta)} =0\,.
}

\item In the second approach, the scalar fields $\chi_{\alpha}$ participate in
the local symmetry transformations and cannot be left out. 
For the abelian case, this realization first appeared in \cite{Alvarez:1993qi} (see also \cite{Hull:2006qs}),
but here we present the generalization to the non-abelian case. 
To our knowledge, this has not appeared in the literature before.\footnote{We
thank F. Rennecke for collaboration on this part.}
The local variations of the action \eqref{action_02} in the second approach read
\eq{
  \label{variantions_01}
  \hat\delta_{\epsilon} X^i = \epsilon^{\alpha} k_{\alpha}^i \,,\hspace{60pt}
  &\hat\delta_{\epsilon} A^{\alpha} = - d\epsilon^{\alpha} - \epsilon^{\beta} A^{\gamma} f_{\beta\gamma}{}^{\alpha}
  \,,\\[1mm]
  &\hat\delta_{\epsilon}\chi_{\alpha} = - \iota_{k_{(\ov \alpha}} v_{\ov \beta)} \epsilon^{\beta}
  - f_{\alpha\beta}{}^{\gamma} 
  \epsilon^{\beta} \chi_{\gamma} 
  \,.
}
However, in this case the constraints are weaker as compared  to \eqref{variantions_33}. 
In particular, they read
\eq{
\label{variations_45}
\mathcal L_{k_{[\ul \alpha}} v_{\ul \beta]} = f_{\alpha\beta}{}^{\gamma} v_{\gamma} \,,
\hspace{40pt}
\iota_{k_{[\ul \alpha}} \op f_{\ul \beta\ul \gamma ]}{}^{\delta} v_{\delta} = \frac{1}{3} \,
\iota_{k_{\alpha}}\iota_{k_{\beta}}\iota_{k_{\gamma}} H \,.
}

\end{enumerate}
Since the local variations \eqref{variantions_01} are in general 
less restrictive as compared to \eqref{variantions_02}, 
in the following we focus on the second approach of implementing the symmetry transformations.


\subsubsection*{Global properties on the world-sheet}

Let us now have a closer examination of the symmetry transformations \eqref{variantions_01},
although we note that the same line of arguments applies to \eqref{variantions_02}.
When varying the action \eqref{action_02}, besides trivial cancellations one is left with 
\eq{
  \label{variation_34}
  \hat\delta_{\epsilon} 
  \widehat{\mathcal S} =-\frac{i}{2\pi\alpha'} \int_{\partial\Sigma}  d\epsilon^{\alpha}
  \wedge (v_{\alpha}+d\chi_{\alpha})
  -\frac{i}{2\pi \alpha'} \int_{\Sigma}  d\epsilon^{\alpha} \wedge dv_{\alpha} \,.
}
In order to show that this variation is vanishing, we assume
that 
$d\epsilon^{\alpha}\wedge (v_{\alpha}+d\chi_{\alpha})$ is globally defined on the 
world-sheet $\partial\Sigma$.
We can then apply Stoke's theorem for the first term in \eqref{variation_34},
canceling  the second term, and leading to 
\raisebox{0pt}[0pt]{$\hat\delta_{\epsilon} \widehat{\mathcal S} = 0$}. 
This assumption follows from a more general requirement, which will be needed later on.
In particular, \label{page_global_ass}
\begin{quote}
We demand that the last line in the gauged action \eqref{action_02} is globally
defined on the world-sheet $\partial\Sigma$, such that Stoke's theorem can be applied.
\end{quote}

This condition imposes some constraints on the fields appearing in the gauged 
world-sheet action \eqref{action_02}, however, a derivation 
of their global properties  from first principles appears to be difficult.
In the case of a single abelian isometry this can be done (see e.g. 
\cite{Rocek:1991ps,Giveon:1993ai,Alvarez:1993qi}), but for the general 
situation we were not able to perform a corresponding analysis. 
We thus leave the global properties of the world-sheet fields unspecified at this point.


\subsubsection*{Generalized geometry}

Let us also give an interpretation of the constraints \eqref{variations_45} in terms of
generalized geometry. 
For the latter, the formal sum of a vector and a one-form is considered to be an element of the 
generalized tangent space which,
with $M$ the target-space manifold,
(locally) takes the form $TM\oplus T^*M$.\footnote{For an introduction to generalized geometry, 
we would like to  refer the reader to the original papers \cite{Hitchin:2004ut} and \cite{Gualtieri:2003dx},
and for instance to \cite{Grana:2008yw} for a discussion in the physics literature.}
The algebraic structure of interest for us is the so-called
$H$-twisted Courant bracket  defined 
as follows
\eq{
  &\bigl[ k_{\alpha} + v_{\alpha} , k_{\beta} + v_{\beta} \bigr]^H_{\rm C} \\[5pt]
  &\hspace{40pt}= 
  \bigl[ k_{\alpha}  , k_{\beta}  \bigr]_{\rm L} 
   + \mathcal L_{k_{\alpha}} v_{\beta} - \mathcal L_{k_{\beta}} v_{\alpha} 
   -\tfrac{1}{2} \, d\bigl( \iota_{k_{\alpha}} v_{\beta} - \iota_{k_{\beta}} v_{\alpha} \bigr)
   -\iota_{k_{\alpha}} \iota_{k_{\beta}} H
  \label{courant_01}
   \,.
}
Using then the relations in \eqref{constraints_35} and \eqref{constraints_351}, and defining
the generalized vectors $K_{\alpha} = k_{\alpha} + v_{\alpha}$, the constraints
\eqref{variations_45} can be written as
\eq{
  \bigl[ K_{\alpha} ,  K_{\beta}  \bigr]^H_{\rm C}
  = f_{\alpha\beta}{}^{\gamma} \op K_{\gamma} \,,
  \hspace{60pt}
  {\rm Nij \op}_{\rm C}^H\bigl( K_{\alpha},  K_{\beta},  K_{\gamma} \bigr) = 0
  \,.
}
The Nijenhuis tensor for the $H$-twisted Courant bracket 
is expressed in terms of the inner product 
$\langle K_{\alpha}, K_{\beta} \rangle = \tfrac{1}{2} (
   \iota_{k_{\alpha}} v_{\beta} + \iota_{k_{\beta}} v_{\alpha} )$
and reads \cite{Gualtieri:2003dx}
\eq{
   {\rm Nij}_{\rm C}^H\bigl(  K_{\alpha},  K_{\beta},  K_{\gamma} \bigr)
   = \bigl\langle \bigl[ K_{[\ul\alpha} ,  K_{\ul \beta}  \bigr]^H_{\rm C} ,
   K_{\ul \gamma ]} \bigr\rangle \,.
}
To summarize, the constraints \eqref{variations_45}
for gauging the non-linear sigma model \eqref{action_01}
by isometries of the target-space manifold are 1) that
the $H$-twisted Courant algebra of generalized vectors $ K_{\alpha} = k_{\alpha}
+ v_{\alpha}$ closes, and 2) that the corresponding Nijenhuis tensor vanishes.


\subsubsection*{Global symmetries of the gauged action}
\label{page_global}

We finally discuss global symmetries of the gauged action.
Suppose that only a subgroup $H\subset G_{\rm iso}$ of the full isometry group $G_{\rm iso}$
has been gauged in \eqref{action_02}. We  denote the Killing vectors corresponding to the gauged
isometry group $H$ by $\{k_{\tilde \alpha}\}$, and we denote the remaining Killing vectors by $\{Z_{\alpha}\}$.
For this setting we find that the gauged action \eqref{action_02} is invariant under global symmetries
parametrized by $Z_{\alpha}$ if
\eq{
  \label{global}
   \bigl[ k_{\tilde\alpha} , Z_{\beta} \bigr]_{\rm L} = 0 
   \hspace{50pt}{\rm and}\hspace{50pt}
   \mathcal L_{Z_{\alpha}} v_{\tilde\beta} = 0\,.
}
Thus, the gauging procedure can break some of the remaining global symmetries in the gauged action.

\vfill


\clearpage
\section{Collective T-duality}
\label{sec_nonab_t}

In this section, we study collective T-duality transformations in detail.
These have been discussed mainly in the context of non-abelian T-duality,
for which some of the main references are 
\cite{delaOssa:1992vc,Giveon:1993ai,Alvarez:1993qi,Sfetsos:1994vz,Alvarez:1994np,Klimcik:1995ux,Lozano:1995jx,Curtright:1996ig}
However, collective T-dualities also include the case of multiple abelian duality
transformations, which have been investigated for instance in \cite{Hull:2006qs}.

As compared to the older references, we approach non-abelian T-duality from a slightly different point of 
view, which for instance makes a particular constraint apparent, and which
does not depend on a gauge-fixing procedure. 
In particular, when following Buscher's procedure \cite{Buscher:1987sk,Buscher:1987qj,Buscher:1985kb} 
of gauging a sigma model and integrating out either the gauge fields or the Lagrange multiplies, 
it is known how to obtain the dual
theory. However, to our knowledge, 
in the non-abelian case it is not known how to recover the original model
without fixing a particular gauge.
Here, we present a mechanism of how the original model can indeed be recovered, 
at least at the classical level, and we discuss the construction of the dual model in the formalism of 
\cite{Plauschinn:2013wta}.


\subsection{Recovering the original model}
\label{sec_recover}

Given the gauged action \eqref{action_02}, one can ask how the original model can 
be recovered. Usually, this is achieved by using the equations of motion for 
the scalar fields $\chi_{\alpha}$,
and for an abelian isometry algebra this has been discussed in
\cite{Rocek:1991ps,Giveon:1993ai,Alvarez:1993qi}
(see also \cite{Kiritsis:1991zt,Kiritsis:1993ju,Giveon:1993ph} for previous as well as
for related work on T-duality transformations),
but for the non-abelian case we are not aware of results in the literature (without fixing a
gauge).


\subsubsection*{Equations of motion for $\chi_{\alpha}$}

We start by determining the equations of motion for the scalar fields $\chi_{\alpha}$. 
For the variation of the action \eqref{action_02} with respect to $\chi_{\alpha}$ we obtain
\eq{
  \delta_{\chi} \widehat{\mathcal S} =+\frac{i}{2\pi\alpha'} \int_{\partial\Sigma} 
  \delta\chi_{\alpha} \op \Bigl( dA^{\alpha} - \tfrac{1}{2}\op f_{\beta\gamma}{}^{\alpha} A^{\beta}\wedge A^{\gamma} \Bigr)\,,
}
from which  we can  read off the equations of motion  as 
\eq{
  \label{eom_01}
  0 = d A^{\alpha} -\tfrac{1}{2}\op f_{\beta\gamma}{}^{\alpha} A^{\beta}\wedge A^{\gamma}  \,.
}


\subsubsection*{Rewriting the action}

We now want to recover the original theory \eqref{action_01} from the gauged version \eqref{action_02}
by employing the equations of motion for $\chi_{\alpha}$.
To this end, let us  define
\eq{
  \label{back_01}
  DX^i = dX^i + k_{\alpha}^i A^{\alpha} \,,
}
and use  Stoke's theorem together with the equation of motion \eqref{eom_01} and the constraints \eqref{variations_45}. After some manipulations we find
\eq{
  \label{action_03}
  \widehat{\mathcal S} =& -\frac{1}{4\pi \alpha'} \int_{\partial\Sigma} 
  \Bigl[ G_{ij} \op DX^i\wedge\star DX^j
  + \alpha'R\, \phi \star 1 \Bigr] \\[2mm]
  &-\frac{i}{2\pi \alpha'} \int_{\Sigma} \tfrac{1}{3!}\, H_{ijk}\op DX^i\wedge DX^j\wedge DX^k
  \,.
}
The structure of this action suggests that in order to obtain the original model, we should perform a 
field redefinition and identify $DX^i$ with the differentials of new coordinates $Y^i$, that is 
$DX^i\to d Y^i$. 
However, in general the one-forms $DX^i$ are not closed, that is
\eq{
  d(DX^i) = \bigl( \partial_m k^i_{\alpha} \bigr) DX^m \wedge A^{\alpha} \,,
}
and therefore such a naive field redefinition would be inconsistent. 
An exception is the case of constant Killing-vector components $\partial_m k^i_{\alpha}=0$, 
corresponding to an abelian isometry algebra, where the simple replacement $DX^i\to d Y^i$ is indeed possible
\cite{Rocek:1991ps,Giveon:1993ai,Alvarez:1993qi}.
For the general case with non-constant Killing vectors, 
a more involved procedure has to be followed. Schematically, it consists of the following steps:
\begin{enumerate}

\item Perform a change of basis of the cotangent space, such that the exterior derivative $d$ acting
on $\{DX^a\}$ in the new basis forms a closed algebra with some structure constants $C_{bc}{}^a$
\eq{
  \label{back_70}
  d (DX^a) = - \frac{1}{2}\op C_{bc}{}^a DX^b\wedge DX^c \,.
}

\item Identify the one-forms $\{DX^a\}$ with vielbeins $E^a = E^a{}_i\op dY^i$,
expressed in terms of new local coordinates $\{Y^i\}$. Note that the vielbeins $\{E^a\}$ 
satisfy the algebra \eqref{back_70}.

\item Perform an inverse change of basis and express the vielbeins $\{E^a\}$ 
in terms of the new differentials $\{dY^i\}$. The action \eqref{action_03} then takes the same form
as the original model \eqref{action_01}.

\end{enumerate}
Note that these steps are simply the generalization from the abelian  to
the non-abelian case. In the following paragraphs, the technical details of this procedure
will be explained; the reader not interested in those  can safely skip to page~\pageref{dual_model}.


\subsubsection*{Change of basis}

Before we begin our discussion, let us impose one technical requirement:
 we demand that the target-space manifold $M$ under consideration has been split 
as 
\eq{
  \label{split}
  M = M_0 \times M_1 \,,
}
where the Killing vectors $\{k_{\alpha}\}$ appearing in the gauged action \eqref{action_02}
form a basis of the tangent space of $M_0$, but are not contained in 
$TM_1$.
Note that the separation \eqref{split} corresponds to choosing so-called {\em adapted coordinates}.
Physically, it means that we perform a T-duality transformation only on $M_0$
and leave $M_1$ unchanged. In the remainder of this section, we  only focus on $M_0$.

In order to perform the field redefinition for a non-abelian isometry algebra, 
let us introduce a new basis for the tangent and co-tangent space
by considering invertible matrices $e^a{}_i=e^a{}_i(X)$ 
with $a,i =1,\ldots,d_0$, and $d_0$ the dimension of $M_0$. 
These matrices do not need to diagonalize the metric,
but in the following we nevertheless refer to them as a vielbein basis.
We then define
\eq{
  \label{back_02}
  e^a = e^a{}_i \op dX^i \,, \hspace{60pt} 
  e_a{} = e_a{}^i\op  \partial_i \,,
}
where $e_a{}^i\equiv (e^{-1})_a{}^i$.
The structure constants  for the dual basis of vector fields $\{e_a\}$ will be denoted by
$C_{ab}{}^c$, and they appear in the  commutator
\eq{
  \label{back_20}
  [e_a,e_b]_{\rm L}= C_{ab}{}^c \op e_c \,.
}
Let us note that by requiring a torsion-free connection, we see that the  one-forms $\{e^a\}$
satisfy the following algebra with respect to the exterior derivative
\eq{
  \label{back_22}
  de^a = -\frac{1}{2} C_{bc}{}^a e^b\wedge e^c\,.
}
We also mention that in regard to this basis the standard notation will be employed, that is
indices are changed from $\{i,j,k,\ldots\}$ to $\{a,b,c,\ldots\}$ by appropriately contracting 
with $e^a{}_i$ or $e^i{}_a$.
Now, the main requirement for the vector fields $\{e_a\}$ defined in \eqref{back_02}
is that they should 
commute with the Killing vector fields $\{k_{\alpha}\}$, that is 
\eq{
  \label{back_03}
    \bigl[k_{\alpha},e_a\bigr]_{\rm L} = 0\,.
}
It is not clear whether a basis of vielbeins satisfying this condition can always be found,
however, in section \ref{sec_sphere} and in appendix \ref{app} we give two explicit examples where this condition is indeed satisfied.


\subsubsection*{Coordinate dependence of the metric, $H$-flux and dilaton}

For the new basis introduced in the previous paragraph we can determine the exterior derivative of 
the one-forms 
$DX^a= e^a{}_i DX^i$. Employing the equation of motion 
shown in \eqref{eom_01}, the algebra \eqref{constraints_351}, and the condition \eqref{back_03},
we find that the one-forms $\{DX^a\}$ form a closed algebra under $d$
\eq{
  \label{back_21}
  d (DX^a) = - \frac{1}{2}\op C_{bc}{}^a DX^b\wedge DX^c \,.
}
Furthermore, using the condition \eqref{back_03} 
together with \eqref{constraints_35} and $dH=0$, we observe that 
the components of the metric and $H$-flux in the vielbein basis satisfy
\eq{
  \label{back_971}
  k_{\alpha}^m \partial_m G_{ab} = 0 \,, \hspace{80pt}
  k_{\alpha}^m \partial_m H_{abc} = 0 \,.
}
Since the Killing vectors $\{k_{\alpha}\}$ span $TM_0$, equations \eqref{back_971}
imply  that these components
are constant on $M_0$. Including then the condition $k_{\alpha}^m\partial_m \phi$
following from \eqref{constraints_35}, in formulas we have that on $M_0$
\eq{
  \label{back_65}
  G_{ab} = {\rm const.} \,, \hspace{50pt}
  H_{abc} = {\rm const.} \,, \hspace{50pt}
  \phi = {\rm const.} \,.
}


\subsubsection*{Recovering the original model}

We are now in the position to show how the original action \eqref{action_01} can be recovered 
from the gauged action \eqref{action_02}. To do so, we first define the one-forms
\eq{
  \label{back_670}
   E^{a} = DX^{a} = e^{a} + k^{a}_{\alpha} A^{\alpha}\,,
}
which by definition satisfy the algebra shown in \eqref{back_21}, that is
\eq{
  \label{back_23}
   dE^a = -\frac{1}{2} C_{bc}{}^a E^b\wedge E^c\,.
}
We observe that this is the same algebra as in \eqref{back_22} which is obeyed by the original 
vielbein one-forms $\{e^a\}$. 
It is therefore clear that a local basis $\{dY^i\}$ of the cotangent space $T^*M_0$ exists, for which 
we can  write
\eq{
  E^a = E^a{}_i \op dY^i \,,
}
with $\{E^a{}_i\}$ invertible matrices.
Now, since the dilaton and the components of the metric and $H$-flux are constant
in the vielbein basis, cf. \eqref{back_65}, we can
rewrite for instance the metric term in the action \eqref{action_03} in the following way
\eq{
  G_{ij}\op  DX^i\wedge\star DX^j = G_{ab}\op  DX^a\wedge\star DX^b
  = G_{ab}\op  E^a\wedge\star E^b 
  = G_{ij}\op  dY^i\wedge\star dY^j \,,
}
where in the last step we performed the inverse change of basis.
An analysis similar to that of the metric can be performed for the $H$-field and dilaton term, so that after 
the above field redefinition we recover from \eqref{action_03} the original action
\eq{
  \mathcal S =& -\frac{1}{4\pi \alpha'} \int_{\partial\Sigma} 
  \Bigl[ G_{ij} \op dY^i\wedge\star dY^j
  + \alpha'R\, \phi \star 1 \Bigr] \\[2mm]
  &-\frac{i}{2\pi \alpha'} \int_{\Sigma} \tfrac{1}{3!}\, H_{ijk}\op dY^i\wedge dY^j\wedge dY^k
  \,.
}
This action may take a different form in the local coordinates $\{Y^i\}$ as compared
to the action in the coordinates $\{X^i\}$.
However, since both of these actions can be expressed in a vielbein basis with the same structure constants,
shown in \eqref{back_22} and \eqref{back_23},
both choices are related by a change of basis.


\subsection{Obtaining the dual model}
\label{dual_model}

Let us now turn to the dual model. As usual, it is obtained by using the equations of motion for 
the gauge fields $A^{\alpha}$ in the gauged action \eqref{action_02}. 
This part of the duality is rather well-understood; here, 
we extend  the formalism of \cite{Plauschinn:2013wta} from the abelian to the
non-abelian case.


\subsubsection*{Equations of motion for $A^{\alpha}$}

We begin by deriving the equations of motion for the gauge fields $A^{\alpha}$ from the
gauged action \eqref{action_02}. Setting to zero the variation with respect to the gauge fields
and solving for $A^{\alpha}$,
we  find
\eq{
  \label{eom_30}
  A^{\alpha} = - \Bigl( \bigl[ \mathcal G - \mathcal D \, \mathcal G^{-1} \mathcal D \bigr]^{-1} \Bigr)^{\alpha\beta} \Bigl(
  \mathds 1 + i \star \mathcal D \, \mathcal G^{-1} \Bigr)_{\beta}^{\;\;\gamma} \bigl( k + i \star \xi \bigr)_{\gamma}\,,
}
where we remind the reader that $\alpha,\beta,\gamma=1,\ldots,N$ label the isometries which have been gauged.
In the expression shown in \eqref{eom_30}, we have employed the notation
\eq{
  \label{back_67}
  \arraycolsep2pt
  \begin{array}{lclclcl}
  \mathcal G_{\alpha\beta} &=& k_{\alpha}^i G_{ij} k^j_{\beta} \,, &\hspace{60pt} &
  \xi_{\alpha} &=& d\chi_{\alpha} + v_{\alpha} \,,
  \\[4mm]
  \mathcal D_{\alpha\beta} &=&  \iota_{k_{[\ul \alpha}} v_{\ul \beta]} + f_{\alpha\beta}{}^{\gamma} 
  \chi_{\gamma}\,,
  &&
  k_{\alpha} & = & k^i_{\alpha} G_{ij} dX^j \,,
  \end{array}
}
and have assumed the matrix $\mathcal G_{\alpha\beta}$ to be invertible. 
In the case of a single Killing vector this corresponds to the usual requirement that
$|k|^2 \neq0$, and in formulas it reads
\eq{
  \label{hvm}
  \det \mathcal G \neq 0 \,.
}
Finally, for later purposes, let us define the symmetric and invertible matrix
\eq{
  \label{back_71}
  \mathcal M = \mathcal G  - \mathcal D \, \mathcal G^{-1} \mathcal D \,.
}


\subsubsection*{Enlarged target-space}

In order to obtain the dual model, we follow the procedure which has been described in detail in
\cite{Plauschinn:2013wta}. 
To do so, we first use the solution \eqref{eom_30} to equations of motion for the gauge fields 
in the gauged action \eqref{action_02}. We then obtain 
\eq{
  \label{action_05}
  \check{\mathcal S} = -\frac{1}{4\pi \alpha'} \int_{\partial\Sigma} 
  \Bigl( \check G
  + \alpha'R\, \phi \star 1 \Bigr)
  -\frac{i}{2\pi \alpha'} \int_{\Sigma} \check H
  \,,
}
where the  tensor fields $\check G$ and  $\check H$ are given by
\eq{
  \label{extended_relations}
  \check G &= G + 
  \binom{k}{\xi}^T \hspace{-2.3pt}\left(\begin{matrix} -\mathcal M^{-1} 
  & -\mathcal M^{-1} \mathcal D \op\mathcal G^{-1} \\ 
  +\mathcal M^{-1} \mathcal D \op\mathcal G^{-1}  & +\mathcal M^{-1} \end{matrix} \right)
  \wedge \star\binom{k}{\xi} \,,
  \\[3mm]
  \check H &= H + 
  \tfrac{1}{2}  \op d\left[  \binom{k}{\xi}^T \left(\begin{matrix} 
   +\mathcal M^{-1} \mathcal D \op\mathcal G^{-1} 
  &+\mathcal M^{-1} \\ 
   -\mathcal M^{-1} & -\mathcal M^{-1} \mathcal D \op\mathcal G^{-1} \end{matrix} \right)
  \wedge \binom{k}{\xi} \right] \,.
}
Here and in the following, matrix multiplication for the indices $\alpha,\beta,\ldots$ is understood.
We observe that these two tensor fields can be interpreted as being defined on an
enlarged $(d_0+N)$-dimensional target space, which is locally
described by the coordinates $\{X^i, \chi_{\alpha}\}$ with $i=1,\ldots, d_0$ and
$\alpha=1,\ldots, N$.
For the enlarged cotangent space, a convenient basis of one-forms is given by
$\{dX^i,\xi_{\alpha}\}$.

As observed in \cite{Plauschinn:2013wta} for the abelian case, the component matrix $\check G_{IJ}$
of the enlarged metric tensor  has
null-eigenvectors. Indeed, consider the following vector in the basis dual to $\{dX^i,\xi_{\beta}\}$ 
\eq{
  \label{back_24}
  \check n_{\alpha}= \left( \begin{array}{c} k^i_{\alpha} \\ 
  \mathcal D_{\alpha\beta}  \end{array}\right),
}
for which we find after a somewhat lengthy computation that
\eq{
  \iota_{\check n_{\alpha}} \check G = 0 \,,
  \hspace{80pt}
  \iota_{\check n_{\alpha}} \check H = 0 \,.
}
Note that the first of these conditions implies that the component matrix $\check G_{IJ}$ 
has $N$ eigenvectors
with vanishing eigenvalue.
We also mention that the vectors \eqref{back_24} are Killing vectors for the enlarged
metric $\check G$ and enlarged field strength $\check H$. In particular, including the result for the dilaton,
we find
\eq{
  \label{iso_10}
  \check{\mathcal L}_{\check n_{\alpha}} \check G = 0\,,\hspace{60pt}
  \check{\mathcal L}_{\check n_{\alpha}} \check H = 0\,,\hspace{60pt}
  \check{\mathcal L}_{\check n_{\alpha}} \phi = 0
  \,.
}


\subsubsection*{Obtaining the dual model}

In order to obtain the dual model from the enlarged target space, we proceed as in the abelian
case. We do not repeat the general discussion of \cite{Plauschinn:2013wta} 
for the  non-abelian case here, 
but only want to outline the main idea. 
\begin{itemize}

\item First, we note that since the metric $\check G$ has $N$ eigenvectors with vanishing eigenvalue,
we can perform a change of basis such that
\eq{
  \renewcommand{\arraystretch}{1.4}
  \arraycolsep8pt
  \check G_{IJ} = \left( \begin{array}{c|c}
  0 & 0 \\ \hline
  0 & \check G_{{\alpha} {\beta}} 
  \end{array} \right) ,
}
with $I,J$ collectively labeling $\{dX^i,\xi_{\alpha}\}$.
As can be verified, the same change of basis results
in vanishing components of the field strength $\check H$ along one or more $dX^i$ directions, that is
\eq{
  \check H_{iJK} = 0 \,.
}
This means, after the change of basis, 
in the action \eqref{action_05} no one-forms $d X^i$ with $i=1,\ldots, d_0$ 
are appearing.

\item Second, the components $\check G_{{\alpha}{\beta}}$ and $H_{{\alpha}{\beta}
{\gamma}}$ as well as the dilaton $\phi$ may still depend on the coordinates $X^{i}$.
However, due to the isometries \eqref{iso_10} of the enlarged target space, we 
may go to a convenient but fixed point in the $X^i$-space. Hence, also the
components do not depend on $X^i$ and we have arrived at the dual model.

\end{itemize}
Note that here we have only outlined the main idea of how the dependence on $\{X^i\}$ and $\{dX^i\}$ 
in the action \eqref{action_05} vanishes. However, in the next two sections
we discuss  explicit examples for this procedure.


\subsubsection*{Remark on isometries of the dual background}

It is well-known that non-abelian T-duality transformations can in general not be inverted.
We do not want to address this question in detail in this paper, but only consider the case 
when  part of the isometry group has been gauged in the action.

Let us therefore recall our discussion from page \pageref{page_global} about the remaining
global symmetries  after the gauging procedure.
There, we saw that only those Killing vectors which satisfy \eqref{global} survive as global isometries
in the gauged theory, in addition to the gauged Killing vectors. Hence, in general 
the isometry group for the dual background is reduced.


\clearpage
\section{Examples I: three-torus}
\label{ex_1_torus}

We now want to illustrate the formalism introduced in the last section with the
example of the three-torus with $H$-flux.
After performing one T-duality transformation, one arrives at the so-called
twisted torus  with vanishing field strength, for which the topology is characterized by a 
geometric flux $f$ \cite{Dasgupta:1999ss,Kachru:2002sk}.
Two successive T-dualities result in a
locally-geometric 
but globally non-geometric background 
which carries a $Q$-flux \cite{Hellerman:2002ax,Shelton:2005cf},
and which is also called a T-fold \cite{Hull:2004in}. 
Finally, three successive T-dualities have been argued to give a
locally non-geometric background carrying so-called $R$-flux 
\cite{Bouwknegt:2004ap,Shelton:2005cf,Ellwood:2006my}.

In this section, we  re-derive these results not using successive but 
{\em collective} T-duality transformations. In section \ref{sec_sphere}, 
we then turn to the example of the
three-sphere, and in appendix \ref{app} the results for the twisted three-torus with $H$-flux have 
been summarized.


\subsubsection*{Setup}

Let us start by introducing some notation. We consider a flat three-torus
with  non-trivial field strength $H$.
The components of the metric tensor in the standard basis of
one-forms $\{dX^1,dX^2,dX^3\}$ are chosen to be of the form
\eq{
  \label{metric_01}
  G_{ij} = \left( \begin{array}{ccc} 
  R_1^2 & 0 & 0 \\ 
  0 & R_2^2 & 0 \\
  0 & 0 & R_3^2
  \end{array} \right),
}  
and the topology is characterized by the identifications $X^i\simeq X^i+\ell_{\rm s}$ for $i=1,2,3$.
The components of the field strength $H=dB$ of the Kalb-Ramond field are taken to be constant,
which, keeping in mind the quantization condition \eqref{quantization}, leads to
\eq{
  \label{ex1_metric_98}
  H = h \, dX^1 \wedge dX^2\wedge dX^3 \,,
  \hspace{60pt}
  h\in \ell_{\rm s}^{-1}\op \mathbb Z \,.
  \hspace{-30pt}
}
The Killing vectors for this configuration 
in the basis $\{\partial_1,\partial_2,\partial_3\}$, dual to the above one-forms, can be chosen as
\eq{
  \label{ex1_killing_02}
  \arraycolsep2pt
  k_{1} = \left( \begin{array}{c} 1 \\ 0 \\ 0 \end{array} \right)  ,\hspace{50pt}
  k_{2} = \left( \begin{array}{c} 0 \\ 1 \\ 0 \end{array} \right)  ,\hspace{50pt}
  k_{3} = \left( \begin{array}{c} 0 \\ 0 \\ 1 \end{array} \right)  ,
}  
which satisfy an abelian algebra, that is
\eq{
  [k_{\alpha},k_{\beta}]_{\rm L} = 0\,.
}  
The one-forms $v_{\alpha}$ corresponding to \eqref{ex1_killing_02} are defined through 
equation \eqref{constraints_35},
and up to exact terms they can be written as
\eq{
  \label{ex_1_forms}
\arraycolsep1.2pt
\renewcommand{\arraystretch}{1.4}
\begin{array}{lclclcr}
  v_{1} &=&  h \op\alpha_1 & X^2 \op dX^3 & - h \op \alpha_2 & X^3 \op dX^2 \,,& \hspace{60pt} 
    \alpha_1 + \alpha_2 = 1 \,, \\
  v_{2} &=&  h \op\beta_1 & X^3 \op dX^1 & - h \op \beta_2 & X^1 \op dX^3 \,,& \hspace{60pt} 
    \beta_1 + \beta_2 = 1 \,, \\
  v_{3} &=&  h \op\gamma_1 & X^1 \op dX^2 & - h \op \gamma_2 & X^2 \op dX^1 \,,& \hspace{60pt} 
    \gamma_1 + \gamma_2 = 1 \,.
\end{array}    
}
Note that here $\alpha_{m}$, $\beta_{m}$ and $\gamma_{m}$ are constants
which parametrize a gauge freedom.
In general these one-forms are not globally
defined on the torus, however, due to the equivalence 
$v_{\alpha} \simeq v_{\alpha} + d \Lambda$ for a function $\Lambda$,
we can define the $v_{\alpha}$  on local charts and 
cover the torus consistently (see for instance \cite{Hull:2006qs} for more details).


\subsubsection*{Constraints on gauging  the sigma model}

As we discussed in section \ref{sec_prelim},  in the presence of a non-vanishing
field strength $H$ there are restrictions on which isometries of the sigma model 
can be gauged, c.f.  equation \eqref{variations_45}.
In the present situation, these imply
\eq{
  \label{ex1_05}
  \iota_{k_{\alpha}}\iota_{k_{\beta}}\iota_{k_{\gamma}} H = 0\,,
}  
so that for the example of the three-torus we can 
distinguish the following cases:
\begin{itemize}

\item For vanishing $H$-flux, 
one, two, or three isometries can be gauged. 
These situations are well-known in the literature,
and so in section \ref{sec_ex1_1} we  discuss briefly only the case 
of gauging all three isometries.

\item For non-vanishing $H$-flux we deduce from \eqref{ex1_05} that 
at most two of the three  isometries can be gauged.
The gauging of only a single isometry is well-known and will be reviewed in section \ref{sec_h_1}.
The situation of gauging two isometries will be discussed in section \ref{sec_h_2}.

\end{itemize}


\subsection[1 T-duality]{One T-duality}
\label{sec_h_1}

We begin by considering one T-duality transformation for the three-torus with non-vanishing $H$-flux.
In the present formalism, this has been analyzed in detail in 
\cite{Plauschinn:2013wta} and so we will be brief here.


\subsubsection*{Gauged action and original model}

For simplicity, let us chose the isometry direction along which we perform the T-duality to
correspond to the Killing vector $k_1=\partial_1$. From the $H$-flux 
\eqref{ex1_metric_98} we deduce the following one-form
\eq{
  v =  h \op\alpha \op X^2 \op dX^3  - h \op (1- \alpha)  X^3 \op dX^2 \,,
}
with $\alpha\in\mathbb R$. The gauged action is  obtained from the general expression shown 
in equation \eqref{action_02} and reads (with the dilaton term omitted)
\eq{
  \widehat{\mathcal S} =&-\frac{1}{2\pi\alpha'} \int_{\partial\Sigma} \;  
  \biggl[ \,\tfrac{1}{2}\op R_1^2\op (dX^1 + A)\wedge\star(dX^1 + A)
  + \sum_{{\mathsf i}=2}^3
  \tfrac{1}{2}\op R_{\mathsf i}^2\op dX^{\mathsf i}\wedge\star dX^{\mathsf i} \;\biggr]
  \\[1mm]
  &-\frac{i}{2\pi \alpha'} \int_{\Sigma} \hspace{8pt} h \, dX^1 \wedge dX^2\wedge dX^3
  \\[1mm]
  &-\frac{i}{2\pi \alpha'} \int_{\partial\Sigma} \hspace{5pt}
  ( v + d\chi)\wedge A \,.
}
The ungauged version is recovered by 
using the equation of motion $dA=0$ as well as Stoke's theorem 
for the last term, which agrees with the general form \eqref{action_03}.
Defining then
$dY^1 = dX^1 + A$, $dY^2=dX^2$ and $dY^3=dX^3$, we arrive at the
original action.


\subsubsection*{Dual model}

In order to obtain the dual theory, we first recall the general formulas shown in equation 
\eqref{back_67}. For a non-vanishing field strength and one Killing vector
we  have
\eq{
  \arraycolsep2pt
  \begin{array}{lclclcl}
  \mathcal G &=& R_1^2 \,, &\hspace{60pt} &
  \xi &=& d\chi +  v \,,
  \\[4mm]
  \mathcal D &=&  0 \,,
  &&
  k & = & R_1^2 \, dX^1 \,,
  \end{array}
}
from which we determine, using \eqref{extended_relations}, 
the metric and field strength of the enlarged target space as follows
\begin{align}
  \label{ex1_76}
  \begin{split}
  \check G 
    &= G - R_1^2\, dX^1\wedge \star dX^1+ \frac{1}{R_1^2}\, \xi\wedge\star \xi \\
  &  =  \frac{1}{R_1^2}\, \xi\wedge\star \xi + 
  R_2^2\, dX^2\wedge\star dX^2 + R_3^2\, dX^3\wedge\star dX^3  \,, 
  \end{split} \\[8pt]
  \label{ex1_77}
  \check H  & = H + d\op \Bigl[\op  dX^1 \wedge \xi\op \Bigr] = 0 \,.
\end{align}
Employing these expressions in the action \eqref{action_05}, we have obtained the dual model (up to 
a transformation of the dilaton). Note that the one-form $\xi$ satisfies
\eq{
  \label{ex1_78}
  d\xi = h\op dX^2\wedge dX^3 \,.
}
Hence, as expected, 
\eqref{ex1_76} and \eqref{ex1_77}, together with \eqref{ex1_78}, describe a twisted three-torus with 
vanishing field strength $\check H=0$ \cite{Dasgupta:1999ss,Kachru:2002sk}.


\subsection[2 T-dualities]{Two T-dualities}
\label{sec_h_2}

Next, we turn to the case of two collective T-dualities for a three-torus with non-vanishing $H$-flux,
and T-dualize along the directions of the Killing vectors $k_1 = \partial_1$
and $k_2=\partial_2$.


\subsubsection*{Gauged action and original model}

In this setting, the one-forms $v_{1}$ and $v_2$ corresponding to  $k_{1}$
and $k_2$ 
are shown in \eqref{ex_1_forms}. However, due to the first condition in \eqref{variations_45}, 
here reading
$\mathcal L_{k_{[\ul \alpha}} v_{\ul \beta]} = 0$,
we find a restriction on the constants $\alpha_m$ and $\beta_m$ in \eqref{ex_1_forms}. 
In particular,
for the one-forms $v_{\alpha}$ we obtain
\eq{
  \label{ex_2_2_19}
  &v_{1} =  h \op\alpha \op X^2 \op dX^3  - h \op (1-\alpha)\op  X^3 \op dX^2 \,,\\[3pt]
  &v_{2} =  h \op(1+\alpha) \op X^3 \op dX^1  + h \op \alpha  \op X^1 \op dX^3 \,,
}
with $\alpha\in\mathbb R$.
Given these expressions, we can  write down the gauged action following from
\eqref{action_02} as
\begin{align}
  \nonumber
  \widehat{\mathcal S} =&-\frac{1}{2\pi\alpha'} \int_{\partial\Sigma} \;  
  \biggl[ \;
  \sum_{{\mathsf i}=1}^2
  \tfrac{1}{2}\op R_{\mathsf i}^2\op \bigl(dX^{\mathsf i} + A^{\mathsf i}\bigr)\wedge\star
  \bigl(dX^{\mathsf i} + A^{\mathsf i}\bigr)
  + 
  \tfrac{1}{2}\op R_3^2\op dX^3\wedge\star dX^3 \;\biggr]
  \\[1mm]
  \nonumber
  &-\frac{i}{2\pi \alpha'} \int_{\Sigma} \hspace{8pt} h \, dX^1 \wedge dX^2\wedge dX^3
  \\[1mm]
  &-\frac{i}{2\pi \alpha'} \int_{\partial\Sigma} \;\biggl[ \;
  \sum_{{\mathsf i}=1}^2
  ( v_{\mathsf i} + d\chi_{\mathsf i})\wedge A^{\mathsf i}
  + h\op X^3  A^1\wedge A^2\;
  \Bigr] \,.
\end{align}
The original ungauged model is again obtained via the procedure discussed in section
\ref{sec_recover}, which in the present case is similar to the example of  one
T-duality.


\subsubsection*{Dual model}

In order to determine the dual model, let us recall equation \eqref{back_67}
and evaluate the there-mentioned quantities. We find
\eq{
  \arraycolsep2pt
  \begin{array}{lclclcl}
  \mathcal G_{\alpha\beta} &=& \left(\begin{matrix} R_1^2 & 0 \\ 0 & R_2^2 \end{matrix} \right) \,, &\hspace{60pt} &
  \xi_{\alpha} &=& \displaystyle \binom{d\chi_{1} + v_{1}}{d\chi_2 + v_2} \,,
  \\[6mm]
  \mathcal D_{\alpha\beta} &=& \left(\begin{matrix} 0 & + h\op X^3 \\ -h\op X^3 & 0 \end{matrix} \right)\,,
  &&
  k_{\alpha} & = & \displaystyle \binom{R_1^2\op dX^1}{R_2^2\op dX^2} \,,
  \end{array}
}
and the matrix $\mathcal M_{\alpha\beta}$ defined in \eqref{back_71} 
takes the following form
\eq{
  \mathcal M_{\alpha\beta} = \left( \begin{matrix} R_1^2 + \bigl[ \frac{h\op X^3}{R_2} \bigr]^2 & 0 \\
  0 & R_2^2 + \bigl[ \frac{h\op X^3}{R_1} \bigr]^2 \end{matrix} \right) .
}
The general formula for the metric of the enlarged target-space was given in
equation \eqref{extended_relations}, which in the basis $\{dX^i, \xi_{\alpha}\}$ becomes
\eq{
  \label{ex1_49}
  \renewcommand{\arraystretch}{1.25}
  \check G_{IJ} = \frac{1}{\rho} \left( \begin{array}{ccc|cc}
  \bigl[ R_1\op h\op X^3 \bigr]^2  & 0 & 0 & 0 & -R_1^2\op h \op X^3 \\
  0 & \bigl[ R_2\op h\op X^3 \bigr]^2 & 0 & +R_2^2\op h \op X^3 & 0 \\
  0 & 0 & \rho\op R_3^2 & 0 & 0 \\ \hline
  0 &  +R_2^2\op h \op X^3 & 0 & R_2^2 & 0 \\
  -R_1^2\op h \op X^3 & 0 & 0 & 0 & R_1^2
  \end{array}\right) ,
}
where for notational convenience we have defined the quantity
\eq{
  \label{ex1_37}
    \rho = R_1^2 R_2^2 + \bigl[h\op X^3\bigr]^2 \,.
}
Next, recall that the matrix \eqref{ex1_49} has eigenvectors with vanishing eigenvalue;
the eigenvectors can therefore be used to perform a change of coordinates. 
Let us consider $\check{\mathsf{G}}_{AB} = ( \mathcal T^{T} \check 
G\, \mathcal T)_{AB}$, where the matrix $\mathcal T$ is given by
\eq{
  \label{ex1_39}
  \renewcommand{\arraystretch}{1.25}
  \mathcal T^I{}_A =  \left( \begin{array}{ccc|cc}
  1 & 0 & 0 & 0 & 0 \\
  0 & 1 & 0 & 0 & 0 \\
  0 & 0 & 1 & 0 & 0 \\ \hline
  0 & -h\op X^3 & 0 & 1 & 0 \\
  + h\op X^3 & 0 & 0 & 0 & 1 
  \end{array}\right) .
}  
Explicitly evaluating the change of basis  we find
\eq{
  \renewcommand{\arraystretch}{1.25}
  \arraycolsep7pt
  \check{\mathsf{G}}_{AB} = ( \mathcal T^{T} \check G\, \mathcal T)_{AB} =
  \frac{1}{\rho} \left( \begin{array}{ccc|cc}
  0  & 0 & 0 & 0 & 0 \\
  0 & 0 & 0 & 0 & 0 \\
  0 & 0 & \rho\op R_3^2 & 0 & 0 \\ \hline
  0 &  0 & 0 & R_2^2 & 0 \\
  0 & 0 & 0 & 0 & R_1^2
  \end{array}\right) .
}
A similar analysis can be carried out for the field strength:
from \eqref{extended_relations} we determine
an expression for $\check H_{IJK}$ 
and we perform the above
change of coordinates, that is
\eq{
  \check{\mathsf H}_{ABC} = \check H_{IJK} \mathcal T^I{}_A\mathcal T^J{}_B\mathcal T^K{}_C\,.
}  
We then find that the only non-vanishing resulting component is
\eq{
  \check{\mathsf H}_{3\xi_{1}\xi_2} = -h\op \frac{R_1^2R_2^2 - \bigl[ h \op X^3\bigr]^2}{\rho^2} \,.
}
Finally, we have to determine how the basis one-forms $\{dX^i\}$ and $\{\xi_{\alpha}\}$ 
transform under the change of basis given by \eqref{ex1_39}.
A short computation leads to 
\eq{
  \mathsf e = \mathcal T^{-1} \left( \begin{array}{c} dX^1 \\ dX^2 \\ dX^3 \\ \xi_1 \\ \xi_2 \end{array}\right)
  = \left( \begin{array}{c} dX^1 \\ dX^2 \\ dX^3 \\ 
  d\bigl( \chi_1 + h\op\alpha X^2X^3\bigr) \\ 
  d\bigl( \chi_2 + h\op\alpha X^1X^3\bigr)
  \end{array}\right) \,,
}
where the free parameter $\alpha\in\mathbb R$ was defined in
\eqref{ex_2_2_19}.
For the dual the model we therefore have the following metric and field strength 
\eq{
  \label{ex1_83}
  &\check{\mathsf G} = \frac{1}{\rho} \,\Bigl[ R_1^2 \, d\tilde X^1\wedge\star d\tilde X^1
  + R_2^2 \, d\tilde X^2\wedge\star d\tilde X^2 \Bigr]
  + R_3^2 \, dX^3\wedge\star dX^3 \,, \\[6pt]
  &\check{\mathsf H} = 
  -h\op \frac{R_1^2R_2^2 - \bigl[ h \op X^3\bigr]^2}{\rho^2} \, d\tilde X^1\wedge d\tilde X^2\wedge dX^3
  \,,
}
with new local coordinates $\tilde X^1=\chi_1 + h\op\alpha X^2X^3$ and 
$\tilde X^2= \chi_2 + h\op\alpha X^1X^3$.
We also remind the reader that the quantity $\rho$ was defined in equation \eqref{ex1_37},
and we observe that the metric and field strength shown in \eqref{ex1_83}
describe the well-known torus-example of a $Q$-flux background \cite{Hellerman:2002ax,Hull:2004in}.


\subsection[3 T-dualities]{Three T-dualities}
\label{sec_ex1_1}

We finally consider  three collective T-dualities for the three-torus.
As explained below equation \eqref{ex1_05}, in this case 
the $H$-flux  has to vanish and so the one-forms $v_{\alpha}$ 
can be chosen to be zero.
The gauged action \eqref{action_02}
becomes
\eq{
  \label{ex_1_02}
  \check{\mathcal S} =&-\frac{1}{2\pi\alpha'} \int_{\partial\Sigma} \;  
  \sum_{\mathsf i=1}^3 \,\Bigl[\, \tfrac{1}{2}\op R_{\mathsf i}^2 \bigl(dX^{\mathsf i} +  A^{\mathsf i}\bigr)
  \wedge\star\bigl(dX^{\mathsf i} +  A^{\mathsf i}\bigr)  
  + i\op d\chi_{\mathsf i} \wedge A^{\mathsf i} \Bigr]\,,
}
and the ungauged action is recovered from \eqref{ex_1_02}
by noting that the equations of motion for $\chi_{\alpha}$ read $d A^{\alpha}=0$. Applying then Stoke's
theorem we observe that the last term in \eqref{ex_1_02} vanishes. 
For the first terms we define new one-forms $dY^I = dX^I + A^I$, and 
therefore recover the original model.


\subsubsection*{Dual model}

To obtain the dual model, we recall our discussion from section \ref{dual_model}. 
For the present setting, the quantities defined in equation \eqref{back_67}
take the following form
\eq{
  \arraycolsep2pt
  \begin{array}{lclclcl}
  \mathcal G_{\alpha\beta} &=& R_{\alpha}^2 \,\delta_{\alpha\beta}\,, &\hspace{60pt} &
  \xi_{\alpha} &=& d\chi_{\alpha}  \,,
  \\[3mm]
  \mathcal D_{\alpha\beta} &=&  0\,,
  &&
  k_{\alpha} & = & R_{\alpha}^2 \delta_{\alpha \op i} \op dX^i  \,.
  \end{array}
}
Using these expressions, we can determine the metric 
of the enlarged target-space from \eqref{extended_relations} as
\eq{
  \check G = G + 
  \binom{k}{d\chi}^T \hspace{-2.3pt}\left(\begin{matrix} -\mathcal G^{-1} 
  & 0 \\ 
  0  & +\mathcal G^{-1} \end{matrix} \right)
  \wedge \star\binom{k}{d\chi} \;=\; \sum_{{\mathsf i}=1}^3 \frac{1}{R_{\mathsf i}^2}\, d\chi_{\mathsf i}
  \wedge\star d\chi_{\mathsf i} \,,
}
and for the field strength we find
\eq{
  \check H =  
  \tfrac{1}{2}  \op d\left[  \binom{k}{d\chi}^T \left(\begin{matrix} 
   0
  &+\mathcal G^{-1} \\ 
   -\mathcal G^{-1} & 0 \end{matrix} \right)
  \wedge \binom{k}{d\chi} \right] = 
  \sum_{{\mathsf i}=1}^3 
  \op d\op \Bigl[ \, dX^{\mathsf i} \wedge d\chi_{\mathsf i} \,\Bigr] =0\,.
}
Using these two results in the action \eqref{action_05}, we see that
it reduces to the dual theory specified by
\eq{
  \check G^{\alpha\beta} = \left( \begin{array}{ccc} 
  \frac{1}{R_1^2} & 0 & 0 \\ 
  0 & \frac{1}{R_2^2} & 0 \\
  0 & 0 & \frac{1}{R_3^2}
  \end{array} \right),
  \hspace{80pt}
  \check H = 0\,,
}  
where for the metric tensor the basis $\{d\chi_{\alpha}\}$ with $\alpha=1,2,3$ has been employed.
Hence, as expected, we find that a collective T-duality 
along all three directions of a three-torus (without $H$-flux)  inverts the
radii.


\subsection{Summary}

To close our discussion of collective T-duality transformations 
for the three-torus with $H$-flux, let us briefly summarize
our results. First, we have seen that the procedure of performing 
{\em collective} T-duality transformations introduced in section~\ref{sec_nonab_t}
leads to the known results in the case of the torus. 
Our discussion in this section therefore serves as a check of that  formalism. 
Second, the examples we have studied  can be summarized as follows:
\begin{itemize}

\item In the case of vanishing field strength $H=0$, a T-duality transformation along any of the
Killing vectors in \eqref{ex1_killing_02}  inverts the corresponding component in the metric.
For three collective T-dualities we have discussed this situation in section \ref{sec_ex1_1}.

\item For non-vanishing field strengths $H\neq0$, one T-duality leads to the so-called twisted
torus. In the present formalism, this has been discussed in detail in \cite{Plauschinn:2013wta},
whose main results we reviewed in section \ref{sec_h_1}.

\item The case of two {\em collective} T-dualities for $H\neq0$ has been discussed in section 
\ref{sec_h_2}. As expected, we arrive at a $Q$-flux background.

\item Finally, due to the requirement \eqref{ex1_05}, we have seen that for a non-vanishing $H$-flux 
three collective T-dualities cannot be performed within the formalism presented in 
section~\ref{sec_nonab_t}.

\end{itemize}
Let us also mention that in appendix \ref{app}, an analysis similar to the three-torus 
has been performed for the
twisted three-torus with $H$-flux. In this case, different variants of a 
 {\em twisted T-fold} are obtained.


\section{Examples II: three-sphere}
\label{sec_sphere}

In this section, we study collective T-duality transformations for  the three-sphere
with $H$-flux.
Some of the results obtained below have partially appeared already in the 
literature; but here we discuss them in a unified manner similar to the example of the three-torus. 
Furthermore, we note that in contrast to the three-torus with $H\neq0$, the three-sphere with 
appropriately adjusted $H$-flux solves the string equations of motion.


\subsubsection*{Setup}

Let us begin  by specifying the setting we will be working in.
For the three-sphere, we choose the round metric in terms of Hopf coordinates which takes the following form
\eq{
  \label{hopf_02}
  ds^2 = R^2 \,\Bigl[ \,\sin^2\eta \,(d\zeta_1)^2
  + \cos^2\eta \,(d\zeta_2)^2
  + (d\eta)^2\, \Bigr]\,,
}
where $\zeta_{1,2}= 0\ldots 2\pi$ and $\eta=0\ldots \pi/2$, and where $R$ denotes the radius of the 
three-sphere. 
We also consider a non-trivial field strength for the Kalb-Ramond field $B$,
\eq{
  \label{ex_2_0}
  H = \frac{h}{2\pi^2}\op \sin \eta \cos\eta \, d\zeta_1\wedge d\zeta_2 \wedge d\eta \,,
}
for which the quantization condition shown in equation \eqref{quantization} implies that
$h \in  \mathbb Z$.
Let us mention that this model solves the string equations of motion for a constant dilaton $\phi_0$, 
hence it is a proper string theory model,
if the field strength $H$ and the radius $R$ of the three-sphere are related as
\eq{
  \label{wzw_relation}
  R^2 = \frac{h}{4\pi^2} \,.
}


\subsubsection*{Killing vectors}

The isometry group of the three-sphere $S^3$ is $O(4)$, and so there are six 
linearly independent Killing vectors.
Employing the basis of vector fields $\{\partial_{\zeta_1},\partial_{\zeta_2},\partial_{\eta}\}$, the
Killing vectors for the metric \eqref{hopf_02} can
be expressed in the following way
\eq{
  \label{hopf_killing}
  \arraycolsep2pt
  \begin{array}{ll@{\hspace{40pt}}ll}
  k_{1} &\displaystyle =\frac{1}{2}\left(\begin{array}{c} +1 \\ -1 \\ 0 \end{array}\right), &
  \tilde{k}_{1} &\displaystyle =\frac{1}{2}\left(\begin{array}{c} +1 \\ +1 \\ 0 \end{array}\right), 
  \\[22.5pt]  
  k_{2} &\displaystyle =\frac{1}{2}\left(\begin{array}{c} -\sin(\zeta_1-\zeta_2) \cot \eta  \\ -\sin(\zeta_1-\zeta_2) \tan\eta 
     \\ \cos(\zeta_1-\zeta_2) \end{array}\right), &
  \tilde{k}_{2} &\displaystyle =\frac{1}{2}\left(\begin{array}{c} +\sin(\zeta_1+\zeta_2) \cot \eta  \\ -\sin(\zeta_1 +\zeta_2) \tan\eta 
     \\ -\cos(\zeta_1+\zeta_2) \end{array}\right), 
  \\[22.5pt]  
  k_{3} &\displaystyle =\frac{1}{2}\left(\begin{array}{c} -\cos(\zeta_1-\zeta_2)\cot \eta  \\ -\cos(\zeta_1-\zeta_2) \tan\eta 
     \\ -\sin(\zeta_1-\zeta_2) \end{array}\right), &
  \tilde{k}_{3} &\displaystyle =\frac{1}{2}\left(\begin{array}{c} +\cos(\zeta_1+ \zeta_2) \cot \eta  \\ -\cos(\zeta_1 +\zeta_2 )\tan\eta 
     \\ +\sin(\zeta_1+\zeta_2) \end{array}\right).
  \end{array}
}
Next, we note that $\mathfrak{so}(4)\cong \mathfrak{su}(2)\times\mathfrak{su}(2)$,
which implies that the above Killing vectors satisfy the following algebra
(with $\alpha,\beta,\gamma\in\{1,2,3\}$ and  $\epsilon_{\alpha\beta\gamma}$ the Levi-Civita symbol)
\eq{
  \label{ex_2_78}
  [ k_{\alpha} , k_{\beta} ]_{\rm L} =  \epsilon_{\alpha\beta}{}^{\gamma} \op 
  k_{\gamma} \,, 
  \hspace{40pt}
  [ k_{\alpha} , \tilde k_{\beta} ]_{\rm L} =  0 \,,
  \hspace{40pt}
  [ \tilde k_{\alpha} , \tilde k_{\beta} ]_{\rm L} =  \epsilon_{\alpha\beta}{}^{\gamma} \op 
  \tilde k_{\gamma} \,.
}
Furthermore, the Killing vectors shown in \eqref{hopf_killing} have constant non-vanishing norm,
corresponding to the fact that they  are 
dual to the invariant one-forms on the three-sphere
\eq{
  | k_{\alpha}|^2 = | \tilde k_{\alpha}|^2 = \frac{R^2}{4} \,.
}


\subsubsection*{Constraints on gauging  the sigma model}

After having introduced our notation, let us now
investigate under which conditions the corresponding non-linear sigma model can
be gauged. These constraints are governed by \eqref{variations_45}, 
however, in order to obtain the dual model we also have to check the
condition \eqref{hvm}.
We consider three different cases:
\begin{itemize}

\item First, gauging a single isometry of the three-sphere 
has been discussed for instance in  \cite{Bouwknegt:2003vb}, and in the present formalism in \cite{Plauschinn:2013wta}.
In this case, the constraint \eqref{variations_45} is always satisfied, and so we
can allow for a non-trivial field strength $H\neq0$. 
Also, since all vectors in \eqref{hopf_killing} 
have constant norm,  the condition \eqref{hvm} is satisfied.

\item Second, for the case of two Killing vectors  we have to choose 
one vector from $\{k_{\alpha}\}$ and one from $\{\tilde k_{\alpha} \}$
in order to obtain a closed algebra.
Because these Killing vectors commute,  the second constraint in \eqref{variations_45} is always 
satisfied.

Without loss of generality, let us then take $\mathsf k_1 = k_1$ and $\mathsf k_2 = \tilde{k}_1$ and determine
the metric $\mathcal G_{\alpha\beta}$ defined in \eqref{back_67}. We find that
\eq{
  \mathcal G_{\alpha\beta} = \frac{R^2}{4} \left( \begin{matrix} 1 & - \cos(2\eta) \\ -\cos(2\eta) & 1 \end{matrix}
  \right) 
  \hspace{20pt}\longrightarrow\hspace{20pt}
  \det \mathcal G = \frac{R^4}{16}\, \sin^2(2\eta) \,,
}
and thus  \eqref{hvm} is not met at the two points $\eta=0$ and $\eta=\pi/2$.

\item Third, the most interesting case is to gauge three isometries. 
Due to the requirement of a closed algebra of Killing vectors, we choose the three
vectors $\{ k_{\alpha}\}$. For those we compute
\eq{
  \mathcal G_{\alpha\beta} = \frac{R^2}{4} \delta_{\alpha\beta} \,,
}
and so the constraint \eqref{hvm} is satisfied. 
However, the conditions  \eqref{variations_45} require a vanishing field strength $H=0$.

\end{itemize}


\subsection[1 T-duality]{One T-duality}

We start with one T-duality for the three-sphere with $H$-flux. 
In the present formalism, this situation
has been analyzed in detail in \cite{Plauschinn:2013wta},
which we review briefly. For the Killing vector, we choose $k=k_1$
from \eqref{hopf_killing}, that is
\eq{
  \label{ex_2_70}
  \arraycolsep3pt
  k = \frac{1}{2} \left(\begin{array}{c} +1 \\ -1 \\ 0 \end{array} \right)
}
in the basis $\{\partial_{\zeta_1},\partial_{\zeta_2},\partial_{\eta}\}$.
The corresponding one-form $v$ 
is determined as
\eq{
  \label{ex_2_7}
  v  = -\frac{h}{8\pi^2} \,\Bigl[ \,
   \alpha_1 \op \zeta_1\op d{\cos}^2\eta - \beta_1 \op{\cos}^2\eta \op d\zeta_1
  + \alpha_2 \op \zeta_2\op d{\cos}^2\eta - \beta_2 \op{\cos}^2\eta \op d\zeta_2 \, \Bigr]\op,  
}
with $\alpha_m+\beta_m=1$.
The gauged action for this setting can be determined from the general form \eqref{action_02}
using the  metric $G$ in \eqref{hopf_02}
together with the one-form $v$ in \eqref{ex_2_7}.
The original model is recovered similarly
to the example of the torus, as the components of $k$ in \eqref{ex_2_70} are constant.
Alternatively, following the discussion in section~\ref{sec_recover},
we note that the relation \eqref{back_03} can be satisfied by 
choosing for $e$ any vector from $\{ k_1, \tilde k_{\alpha}\}$ shown in 
\eqref{hopf_killing}, which then obeys $[ k , e ]_{\rm L}=0$.


\subsubsection*{Dual model}

To obtain the dual model, we start by determining the quantities in \eqref{back_67} 
for the present setting:
\eq{
  \arraycolsep2pt
  \begin{array}{lclclcl}
  \mathcal G &=& \frac{R^2}{4} \,, &\hspace{60pt} &
  \xi &=& d\chi + v \,,
  \\[4mm]
  \mathcal D &=&  0 \,,
  &&
  k & = & \frac{R^2}{2} \Bigl( \sin^2\eta \op d\zeta_1 - \cos^2\eta \op d\zeta_2 \Bigr)\,;
  \end{array}
}
the matrix  $\mathcal M$ is given by $\mathcal M = \mathcal G$. The metric and 
field strength of the enlarged target space appearing in the action \eqref{action_05}
are determined by the general expressions \eqref{extended_relations}, which in the present 
case become
\eq{
  \label{ex_2_1}
  &\check G = R^2 \left[ d\eta\wedge\star d\eta +\frac{1}{4}\sin^2(2\eta) \,(d\zeta_1+d\zeta_2)
  \wedge \star (d\zeta_1+d\zeta_2) \right]
    + \frac{4}{R^2}\,\xi\wedge \star \xi \,, \\[8pt]
  &\check H = 2\sin(2\eta) \op (d\zeta_1+d\zeta_2)\wedge d\eta\wedge \xi \,.
}
If we now make the redefinitions $\tilde\eta = 2 \eta$ and $\tilde\zeta=\zeta_1+\zeta_2$, 
we can express the above metric and field strength as
\eq{
  \label{ex_2_5}
  &\check{\mathsf G} = \frac{R^2}{4} \,\Bigl[\op  d\tilde\eta\wedge\star d \tilde\eta +\sin^2\tilde\eta \,d\tilde\zeta
  \wedge\star d\tilde\zeta \Bigr]
    + \frac{4}{R^2}\,\xi\wedge \star \xi \,, \\[8pt]
  &\check{\mathsf H} = \sin\tilde\eta\, \op d\tilde\zeta\wedge d\tilde \eta\wedge \xi \,.
}  
Noting then furthermore that 
\eq{
  \label{ex_2_2}
  d\xi = - \frac{h}{16\pi^2}\,\sin\tilde\eta\, d\tilde\eta \wedge d\tilde\zeta \,,
}
we conclude that the metric $\check {\mathsf G}$ in \eqref{ex_2_5} corresponds to 
a circle of radius $\frac{2}{R}$ which 
is fibered over a round two-sphere of radius $\frac{R}{2}$,
with the twisting characterized by \eqref{ex_2_2} \cite{Alvarez:1993qi,Bouwknegt:2003vb}
(see also \cite{Israel:2004vv,Orlando:2010ay} for related work).
Note that the dual model solves again the string equations of motion for a constant dilaton.


\subsection[2 T-dualities]{Two T-dualities}

Next, we turn to the three-sphere with non-trivial $H$-flux \eqref{ex_2_0}
and  perform two duality transformations along the Killing vectors 
\eq{
  \arraycolsep3pt
  \mathsf k_1 = k_1 = \frac{1}{2} \left(\begin{array}{c} +1 \\ -1 \\ 0 \end{array} \right) \,, 
  \hspace{60pt}
  \mathsf k_2 = \tilde{k}_1 = \frac{1}{2} \left(\begin{array}{c} +1 \\ +1 \\ 0 \end{array} \right) ,
}
which are written in the basis $\{\partial_{\zeta_1},\partial_{\zeta_2},\partial_{\eta}\}$.
The corresponding one-forms are again specified by the second equation in \eqref{constraints_35},
and take the form
\eq{
  \mathsf v_1 & =  \frac{h}{8\pi^2} \,\Bigl[ \,
  - \alpha_1 \op \zeta_1\op d{\cos}^2\eta + \beta_1 \op{\cos}^2\eta \op d\zeta_1
  - \alpha_2 \op \zeta_2\op d{\cos}^2\eta + \beta_2 \op{\cos}^2\eta \op d\zeta_2 \, \Bigr]\op, \\
  \mathsf v_2  &= \frac{h}{8\pi^2} \,\Bigl[ \,
  + \alpha_3 \op \zeta_1\op d{\cos}^2\eta - \beta_3 \op{\cos}^2\eta \op d\zeta_1
  - \alpha_4 \op \zeta_2\op d{\cos}^2\eta + \beta_4 \op{\cos}^2\eta \op d\zeta_2 \, \Bigr]\op.
}
However, in order to satisfy the constraint \eqref{variations_45}, the constants $\alpha_m$ and $\beta_m$
have to be restricted as $ \beta_1+\beta_2+\beta_3+\beta_4 = 4$.
The original model can be recovered along the lines discussed in section~\ref{sec_recover}
by choosing vector fields $\{e_a\} = \{ k_1, \tilde k_1\}$, 
which  satisfy the relation \eqref{back_03}.


\subsubsection*{Dual model}

In order to determine the dual model, we first compute the quantities shown in \eqref{back_67}
for the present example. From the above data we obtain
\eq{
  \label{ex_2_76}
  \arraycolsep2pt
  \begin{array}{@{}lclclcl@{}}
  \mathcal G_{\alpha\beta} &=& 
  \displaystyle \frac{R^2}{4} \left( \begin{matrix} 1 & - \cos(2\eta) \\ -\cos(2\eta) & 1 \end{matrix}
  \right)\! , &\hspace{15pt} &
  \xi_{\alpha} &=& d\chi_{\alpha} + \mathsf v_{\alpha} \,,
  \\[6mm]
  \mathcal D_{12} &=&  \displaystyle -\frac{h}{8\pi^2} \, \cos^2\eta \,,
  &&
  k_{\alpha} & = & \displaystyle \frac{R^2}{2} 
  \left(\begin{array}{c}
    \sin^2\eta \op d\zeta_1 - \cos^2\eta \op d\zeta_2 \\
    \sin^2\eta \op d\zeta_1 + \cos^2\eta \op d\zeta_2 
  \end{array}\right)\!.
  \end{array}
}
The general form of the dual world-sheet action is again \eqref{action_05}, 
where the corresponding metric and field strength are determined by \eqref{extended_relations}.
Employing \eqref{ex_2_76}, we find 
a rather complicated expression for the enlarged metric, which we do not present here.
However, after a change of basis characterized by
\eq{
  \label{change_3}
  \renewcommand{\arraystretch}{1.25}
  \mathcal T^I{}_A =  \left( \begin{array}{ccc|cc}
  +\frac{1}{2} & +\frac{1}{2} & 0 & 0 & 0 \\
  -\frac{1}{2} & +\frac{1}{2} & 0 & 0 & 0 \\
  0 & 0 & 1 & 0 & 0 \\ \hline
  0 & -\mathcal D_{12} & 0 & +\frac{1}{2} & +\frac{1}{2} \\
  +\mathcal D_{12} & 0 & 0 & +\frac{1}{2} & -\frac{1}{2} 
  \end{array}\right) ,
}
we obtain for the metric tensor in the new basis 
\eq{
  \label{ex_2_018}
  \renewcommand{\arraystretch}{1.25}
  \arraycolsep7pt
  \check{\mathsf{G}}_{AB} = ( \mathcal T^{T} \check G\, \mathcal T)_{AB} =
  \left( \begin{array}{ccc|cc}
  0  & 0 & 0 & 0 & 0 \\
  0 & 0 & 0 & 0 & 0 \\
  0 & 0 &  R^2 & 0 & 0 \\ \hline
  0 &  0 & 0 & \check{\mathsf G}_{11}& 0 \\
  0 & 0 & 0 & 0 &  \check{\mathsf G}_{22}
  \end{array}\right) ,
}
where the two components $ \check{\mathsf G}_{11}$ and $ \check{\mathsf G}_{22}$ are 
given by
\eq{
 &\check{\mathsf G}_{11}=\frac{1}{R^2} \left[ \sin^2\eta + \left( \frac{h}{4\pi^2 R^2}\right)^2 \cos^2\eta\right]^{-1}
 \,, \\
  &\check{\mathsf G}_{22}=\frac{1}{R^2} \left[ \cos^2\eta + \left( \frac{h}{4\pi^2 R^2}\right)^2 
  \frac{\cos^4\eta}{\sin^2\eta} \right]^{-1}  \,.
}
The one forms in the transformed basis take the following general form
\eq{
 \mathsf e^A = \mathcal T^{-1} \left( \begin{array}{c} d\zeta_1 \\ d\zeta_2 \\ d\eta \\ \xi_1 \\ \xi_2 \end{array}\right)
 \,,
}
and we note that $\mathsf e^{\xi_1}$ and $\mathsf e^{\xi_2}$ are exact, so we can 
introduce new coordinates
$\tilde\zeta_1$ and $\tilde\zeta_2$ via $\mathsf e^{\xi_1}= d\tilde\zeta_1$ and $\mathsf e^{\xi_2}= d\tilde\zeta_2$.
A similar analysis can be performed for the dual field strength: using the expression shown
in \eqref{extended_relations} and performing the change of basis characterized by
\eqref{change_3}, we find that the only non-vanishing component of $\check{\mathsf H}_{ABC}$
reads
\eq{
  \label{ex_2_019}
  \check{\mathsf H}_{\eta\xi_1\xi_2} = 
  -8\op h\op \pi^2 \bigl( h^2 - 16\pi^2 R^4 \bigr)
  \,\frac{\sin\eta\cos\eta}{\left[ 16\pi^2 R^4 \sin^2\eta + h^2 \cos^2\eta\right]^2} \,.
}


\subsubsection*{Summary and discussion}

The expressions for the components of the dual metric and field strength were given in  
equations \eqref{ex_2_018} and \eqref{ex_2_019}, which we summarize as
\begin{align}
  \nonumber
  &\check{\mathsf G} = R^2 \op d\eta\wedge\star d\eta +
  \frac{1}{R^2} \, \frac{d\tilde\zeta_1\wedge\star d\tilde\zeta_1}
  { \sin^2\eta + \left[ \frac{h}{4\pi^2 R^2}\right]^2 \cos^2\eta}
  + \frac{1}{R^2} \, \frac{d\tilde\zeta_2\wedge\star d\tilde\zeta_2}{
  \cos^2\eta + \left( \frac{h}{4\pi^2 R^2}\right)^2 
  \frac{\cos^4\eta}{\sin^2\eta}} \,, \\[8pt]
  \label{ugly}
  &\check{\mathsf H} = 
  -8\op h\op \pi^2 \bigl( h^2 - 16\pi^4 R^4 \bigr)
  \,\frac{\sin\eta\cos\eta}{\left[ 16\pi^2 R^4 \sin^2\eta + h^2 \cos^2\eta\right]^2} \, 
  d\eta\wedge d\tilde\zeta_1\wedge d\tilde\zeta_2
  \,.
\end{align}
These formulas are rather complicated, and it appears to be difficult to extract 
properties of the dual space. However, if we use the condition  \eqref{wzw_relation}
for solving the string equations of motion of the original model, the above formulas simplify
considerably. In particular, we find
\eq{
  \label{ex_2_2_07}
  &\ov{\mathsf G} = R^2 \op d\eta\wedge\star d\eta +
  \frac{1}{R^2}\, \Bigl[  d\tilde\zeta_1\wedge\star d\tilde\zeta_1
  + \tan^2 \eta\, d\tilde\zeta_2\wedge\star d\tilde\zeta_2  \Bigr]\,, \\[8pt]
  &\ov{\mathsf H} = 0
  \,,
}
which describes a non-compact but geometric background. This is in 
contrast to the example of the three-torus with $H$-flux discussed in section~\ref{sec_h_2}, where after two T-dualities
a non-geometric $Q$-flux background was obtained.

Let us also note that the dual configuration \eqref{ex_2_2_07} solves again the string equations of motion 
if we transform 
the dilaton via the standard relation of
the Buscher rules \cite{Buscher:1987sk,Buscher:1987qj,Buscher:1985kb} as
\eq{
  \ov\phi = -\log\bigl( R^2 \cos\eta \bigr) + \phi_0 \,.
}
Note furthermore, this backgrounds is related to Witten's black 
hole \cite{Witten:1991yr}, that is the group manifold
$SL(2,\mathbb R)/U(1)$, via analytic continuation.


\subsection[3 T-dualities]{Three T-dualities}

We finally consider the situation of gauging three (non-abelian) isometries of the three-sphere. 
As explained in the beginning of this
section, in this case the constraints in \eqref{variations_45} require a vanishing
field strength $H=0$. Thus, we have 
\eq{
  H = 0 \hspace{40pt}\longrightarrow\hspace{40pt}
  v_{\alpha} = 0\,.
}
For the Killing vectors, we can choose either of the sets $\{k_{\alpha}\}$ or $\{\tilde k_{\alpha}\}$; 
for definiteness we consider the first in the following.


\subsubsection*{Gauged action and original model}

The gauged action can again be inferred from the general form shown in equation 
\eqref{action_02}.  Using
coordinates $\{X^1,X^2,X^3\}=\{\zeta_1,\zeta_2,\eta\}$, we find
\eq{
  \widehat{\mathcal S} =&-\frac{1}{2\pi\alpha'} \int_{\partial\Sigma} \;  
  \tfrac{1}{2}\op G_{ij}  (dX^i + k^i_{\alpha} A^{\alpha})\wedge\star(dX^j + k^j_{\beta} A^{\beta})  
  \\[1mm]
  &-\frac{i}{2\pi \alpha'} \int_{\partial\Sigma} \:\Bigl[ \;
   d\chi_{\alpha}\wedge A^{\alpha}
  + \tfrac{1}{2}\op f_{\alpha\beta}{}^{\gamma} \chi_{\gamma} \, A^{\alpha}\wedge A^{\beta}\;
  \Bigr] \,,
}
where now the gauge fields are non-abelian. To recover the original ungauged model,
we use the equations of motion \eqref{eom_01} for $A^{\alpha}$ and rewrite the action
as in section \ref{sec_recover}. In particular, from \eqref{action_03} we obtain
\eq{
  \widehat{\mathcal S} = -\frac{1}{4\pi \alpha'} \int_{\partial\Sigma} 
  G_{ij} \op DX^i\wedge\star DX^j
  \,,
}
where $  DX^i = dX^i + k_{\alpha}^i A^{\alpha} $.
However, we note  that $d(DX^i)\neq0$, and so we can not make the 
replacements $DX^i\to dY^i$ as before. 
The way to proceed has been described in section \ref{sec_recover}.
We first need to find a set of vector fields $\{e_a\}$ which commute with $\{k_{\alpha}\}$
and thus satisfy equation \eqref{back_03}. For the three-sphere 
we have an obvious candidate, namely $\{\tilde{k}_{\alpha}\}$,
\eq{
  \arraycolsep4pt
  e_a{}^i = \tilde{k}_{\alpha}{}^i = 
  \scalemath{0.95}{
  \frac{1}{2}
  \left( \begin{array}{ccc}
  1 & 1 & 0 \\
  +\sin(\zeta_1+\zeta_2) \cot \eta & - \sin(\zeta_1+\zeta_2)\tan\eta & - \cos(\zeta_1+\zeta_2) \\
  +\cos(\zeta_1+\zeta_2) \cot \eta & - \cos(\zeta_1+\zeta_2)\tan\eta & + \sin(\zeta_1+\zeta_2)
  \end{array}\right)
  }.
}
The metric \eqref{hopf_02} can then be transformed via
\eq{
  \label{metric_sphere_new}
  G_{ab}  = e_a{}^i \op G_{ij}\op (e^T)^j{}_b = \frac{R^2}{4} \, \delta_{ab} \,,
}
and for $DX^a$ in the new basis we compute
\eq{
  d(DX^a) = -\frac{1}{2}\, \epsilon^a{}_{bc} DX^b\wedge DX^c \,.
}
Hence, the one-forms $\{DX^a\}$ behave like a non-holonomic basis of the co-tangent space. Since
the corresponding metric \eqref{metric_sphere_new} is constant, we can define
new vielbeins $E^a=DX^a$, and express them in a local basis $dY^i$ as
\eq{
  E^a = e^a{}_i \op dY^i \,,
}
where we also performed the obvious relabeling $X^i \to Y^i$ in the matrix $e^a{}_i$. Using 
this form, we then arrive at the original ungauged action
\eq{
  \mathcal S = -\frac{1}{4\pi \alpha'} \int_{\partial\Sigma} 
  R^2 \,\Bigl[ \,\sin^2\eta \,d\zeta_1\wedge\star d\zeta_1
  + \cos^2\eta \,d\zeta_2\wedge\star d\zeta_2
  + d\eta\wedge\star d\eta\, \Bigr]
  \,.
}


\subsubsection*{Dual model}

In order to determine the dual model, we fist specify the quantities in equation \eqref{back_67} as follows
\eq{
  \arraycolsep2pt
  \begin{array}{lclclcl}
  \mathcal G_{\alpha\beta} &=&  \frac{R^2}{4} \delta_{\alpha\beta} \,, &\hspace{60pt} &
  \xi_{\alpha} &=& d\chi_{\alpha}  \,,
  \\[4mm]
  \mathcal D_{\alpha\beta} &=&  \epsilon_{\alpha\beta}{}^{\gamma} \chi_{\gamma}\,,
  &&
  k_{\alpha} & = & k^i_{\alpha} G_{ij} dX^j \,,
  \end{array}
}
where we did not spell out the expression for the one-forms $k_{\alpha}$ corresponding to the
Killing vectors. Using then the general formulas shown in \eqref{extended_relations},
the metric and field strength of the enlarged target-space can be determined.
These expressions become quite involved, and so we only display the quantities 
after a change of basis given by the null-eigenvectors \eqref{back_24} has been performed
and after a field redefinition. In a basis $\{d\tilde\chi_1,d\tilde\chi_2,d\tilde\chi_3\}$
we obtain
\eq{
  \renewcommand{\arraystretch}{1.3}
  &\ov{\mathsf G}_{\alpha\beta} = 
  \frac{4}{R^2} \, \frac{1}{\frac{R^4}{16} + \tilde\chi_1^2+\tilde\chi_2^2+\tilde\chi_3^2}
  \left(
  \begin{array}{ccc}
  \frac{R^4}{16} + \tilde\chi_1^2 & \tilde\chi_1 \tilde\chi_2 & \tilde\chi_1 \tilde\chi_3 \\
  \tilde\chi_1 \tilde\chi_2 & \frac{R^4}{16} + \tilde\chi_2^2 & \tilde\chi_2 \tilde\chi_3 \\
  \tilde\chi_1 \tilde\chi_3 & \tilde\chi_2 \tilde\chi_3 & \frac{R^4}{16} +  \tilde\chi_3^2 \\  
  \end{array}
  \right) , \\[10pt]
  & \ov{\mathsf H}_{123}= 
  16\, \frac{3\frac{R^4}{16} + \tilde\chi_1^2+\tilde\chi_2^2+\tilde\chi_3^2}
  {\frac{R^4}{16} + \tilde\chi_1^2+\tilde\chi_2^2+\tilde\chi_3^2}
   \,.
}  
Performing now a further change to spherical coordinates $\{\rho,\phi_1,\phi_2\}$
with $\rho\geq0$ and $\phi_{1,2}=0,\ldots, 2\pi$, we find
\eq{
  \label{ex_2_10}
  &\ov{\mathsf G} = \frac{4}{R^2}\, d\rho\wedge\star d\rho + \frac{R^2}{4}\, \frac{\rho^2}{\rho^2 + \frac{R^4}{16}}
  \,\Bigl[ d\phi_1\wedge\star d\phi_1 
  + \sin^2(\phi_1) \,d\phi_2\wedge\star d\phi_2
  \Bigr] \,, \\[5pt]
  & \ov{\mathsf H}= \frac{\rho^2}{\left(\rho^2 + \frac{R^4}{16}\right)^2}
  \left[\rho^2 + 3\,\frac{R^4}{16}\right] \sin(\phi_1) \:d\rho\wedge d\phi_1 \wedge d\phi_2 
  \,.
}
This configuration can be interpreted as a two-sphere (parametrized by $\phi_1$ and $\phi_2$)
whose radius depends on the ray-variable $\rho$. 
(The same result has been obtained in \cite{Alvarez:1993qi} and \cite{Curtright:1996ig}, 
and related expressions can be found in \cite{Itsios:2012dc})
Note that the volume of the two-sphere
as well as the $H$-flux
vanish at $\rho=0$, but stay finite in the limit $\rho\to\infty$.


\subsection{Summary}

In this section we have considered collective T-duality transformations for the three-sphere
with $H$-flux. One of the features of this background is that it solves the 
string equations of motion if the flux is adjusted properly, c.f. \eqref{wzw_relation}.
The main purpose of studying this example was to investigate whether results similar to the
three-torus with $H$-flux can be obtained. 
\begin{itemize}

\item After a single T-duality for the three-sphere with $H$-flux, we arrived at the background
of a circle  fibered over a two-sphere. This is a well-defined geometric background with 
geometric flux, which agrees with the result for the torus obtained in section \ref{sec_h_1}.

\item After two collective T-dualities for the three-sphere we obtained at a rather complicated-looking
background, shown in equation \eqref{ugly}. However, when imposing the condition \eqref{wzw_relation}
for the original model to be  conformal, the background simplified considerably. 
In particular, despite being non-compact, the dual background is {\em geometric}.
This is in contrast to our discussion in section \ref{sec_h_2}, where two T-dualities for the three-torus 
lead to a {\em non-geometric} background.

\item Finally, for three collective T-duality transformations we  found that the $H$-flux has to
vanish. The corresponding dual background shown in equation \eqref{ex_2_10} is again geometric but
non-compact.

\end{itemize}


\section{Summary and conclusions}
\label{sec_summary}

In this paper, we have studied T-duality transformations along one, two, and three
directions of isometry for the three-sphere with $H$-flux. The question we
wanted to answer was, whether after two T-dualities a non-geometric
$Q$-flux background similarly to the example of the three-torus appears.

\bigskip
In order to perform the duality transformations, in section \ref{sec_nonab_t} 
we have developed a novel formalism for collective, and in general 
non-abelian, T-duality. Our approach is different compared to the known
literature, as we do not rely on a gauge fixing procedure nor on the specific
structure of Wess-Zumino-Witten models. Furthermore,
we derived a constraint, shown in equation \eqref{variations_45}, which restricts
the allowed transformations in the case of non-vanishing $H$-flux.
For the three-torus and three-sphere this implied that for $H\neq0$ at most
two T-dualities can be performed.

In section~\ref{ex_1_torus} we illustrated our formalism with the example
of the three-torus and reproduced the known results; this analysis served
as a check of our procedure. In addition, in appendix~\ref{app} we studied 
collective T-duality transformations for the twisted torus with $H$-flux, for which 
we found a new {\em twisted T-fold} background.

In section~\ref{sec_sphere} we  investigated collective T-duality transformations for
the three-sphere with $H$-flux. In contrast to the torus, this background solves 
the string equations of motion if the flux is properly adjusted.
For one T-duality, we reproduced the known result,
namely the dual background is a circle fibered over a two-sphere. In view of the duality chain \eqref{t_chain},
this configuration would correspond to a geometric-flux background.
After applying two collective T-dualities, we obtained a rather complicated background,
which resembled the form of the torus T-fold. However, if the radius of the three-sphere
is appropriately related to the $H$-flux, making the original model conformal, the 
dual background simplified considerably. In particular, one obtains a two-sphere 
fibered over a line segment, which is a geometric but non-compact space.
Finally, as mentioned above, for three T-dualities the restrictions \eqref{variations_45} 
require a vanishing $H$-flux. We therefore chose $H=0$ and obtained
after a non-abelian T-duality transformation a two-sphere fibered over
a ray.

\bigskip
Let us  compare our results for two collective T-duality transformations on the three-torus
and on the three-sphere with $H$-flux. For the torus we reviewed that one obtains a non-geometric
$Q$-flux background, or more generally a T-fold. Note however, the torus with $H\neq0$ does not solve
the string equations of motion and therefore is, strictly speaking, not a proper string background.
For the three-sphere, without requiring the model to be conformal, 
we found a background of a form similar to the torus T-fold.
But, after requiring the original model to solve the string equations of motion, the dual 
background simplified. In particular, the dual space is geometric but non-compact.

Our findings in this paper therefore challenge the simple picture of T-duality transformations
shown in \eqref{t_chain}. Namely, applying two T-duality transformations to a geometric
background with $H$-flux  does not necessarily lead to a non-geometric 
$Q$-flux background. However, we also want to emphasize that 
the two examples studied in this paper have drawbacks:
the torus example does not solve the string equations of motion, and the three-sphere leads 
to a non-compact background. We therefore cannot draw general conclusions about the origin
of non-geometry, but have to consider further examples in the future.


\vskip3em
\subsubsection*{Acknowledgements}

We would like to thank F. Rennecke for collaboration at an early stage of the project;
and we thank I. Bakas, R. Blumenhagen and D. L\"ust for useful comments.
We also thank the Max-Planck-Institute for Physics in Munich for hospitality, where part of
this work has been done.
The research of the author is supported by the 
MIUR grant FIRB RBFR10QS5J, and by the COST Action MP1210.


\clearpage
\begin{appendix}

\section{Examples III: twisted three-torus}
\label{app}

As a generalization of the three-torus with $H$-flux, in this appendix we discuss the twisted three-torus
found in section~\ref{sec_h_1} together with a non-vanishing $H$-flux.


\subsubsection*{Setup}

The components of the metric tensor of the twisted three-torus in a coordinate basis 
$\{dX^1,dX^2,dX^3\}$ are chosen as
\eq{
  \label{metric_05}
  \renewcommand{\arraystretch}{1.3}
  G_{ij} = \left( \begin{array}{ccc} 
  R_1^2 & 0 & R_1^2 \op f X^2  \\ 
  0 & R_2^2  & 0 \\
  R_1^2 \op f X^2 & 0 & R_3^2 + R_1^2 \left[ f X^2 \right]^2
  \end{array} \right),
}  
where $f$ denotes the geometric flux, and we allow for a non-vanishing field strength of the
Kalb-Ramond field
\eq{
  \label{ex2_h-field}
  H = h \, dX^1\wedge dX^2\wedge dX^3 \,,
  \hspace{80pt}
  h \in \ell_{\rm s}^{-1}\op \mathbb Z \,.
}  
The Killing vectors for the above metric in the basis
$\{\partial_1,\partial_2,\partial_3\}$ are given by
\eq{
  \arraycolsep2pt
  k_{1} = \left( \begin{array}{c} 1 \\ 0 \\ 0 \end{array} \right)  ,\hspace{60pt}
  k_{2} = \left( \begin{array}{c} - f\op X^3 \\ 1 \\ 0 \end{array} \right)  ,\hspace{60pt}
  k_{3} = \left( \begin{array}{c}  0 \\ 0 \\ 1 \end{array} \right)  ,
}  
which satisfy a non-abelian isometry algebra with commutation relations
\eq{
\bigl[ k_1,k_2 \bigr]_{\rm L} = 0\,, \hspace{50pt}
\bigl[ k_2,k_3 \bigr]_{\rm L} = f\op k_1\,,\hspace{50pt}
\bigl[ k_3,k_1 \bigr]_{\rm L} = 0 \,.
}
Furthermore, the topology of the twisted torus is specified by the identifications
\eq{
  \arraycolsep2pt
  \renewcommand{\arraystretch}{1.25}
  \begin{array}{c@{\hspace{15pt}}lcl@{\hspace{20pt}}lcl}
  1) & X^1 &\rightarrow&X^1 +\ell_{\rm s} \,, \\
  2) & X^2 &\rightarrow&X^2 +\ell_{\rm s} \,, \\
  3) & X^3 &\rightarrow&X^3 +\ell_{\rm s} \,, & X^1 &\rightarrow&X^1 + \ell_{\rm s}\op  fX^2 \,.
  \end{array}
}


\subsubsection*{One T-duality}

As it is well-known \cite{Bouwknegt:2003vb,Bouwknegt:2003wp,Bouwknegt:2003zg},
a single T-duality along the Killing vector $k_1$ results  in a twisted torus with the
replacements
\eq{
  \label{ex2_exchange}
  f \;\longleftrightarrow \;h \,, \hspace{80pt}
  R_1  \;\longrightarrow \; \frac{1}{R_1} \,.
}
However, a T-duality along the Killing vectors $k_2$ or  $k_3$ leads to a  twisted T-fold. 
More concretely, after performing a T-duality
transformation along the Killing vector $k_2$, we find for the dual metric and $H$-field the expressions
\eq{
  \label{dual_master_1}
  &\check G = \frac{1}{1+\bigl[\op \frac{R_1}{R_2} \op f\op  X^3 \bigr]^2}
  \left( R_1^2 \op dX^1 \wedge\star dX^1 + \frac{1}{R_2^2} \op \xi\wedge\star\xi \right)
  + R_3^2 \op dX^3\wedge\star dX^3 \,, \\[8pt]
  &\check H = -f\, \frac{R_1^2}{R_2^2} \: 
  \frac{1-\bigl[\op \frac{R_1}{R_2} \op f\op  X^3 \bigr]^2}
  {\left(1+\bigl[\op \frac{R_1}{R_2} \op f\op  X^3 \bigr]^2\right)^2} \:dX^1 \wedge \xi \wedge dX^3\,,
}
where the one form $\xi$ is not closed,
\eq{
  \label{dual_master_2}
  d \xi = - h\, dX^1 \wedge dX^3 \,.
}
The result for a T-duality along $k_3$ leads to the same expression but with 
the replacements $X^2 \leftrightarrow -X^3$ and $R_2 \leftrightarrow R_3$.


\subsubsection*{Two T-dualities}

When performing two collective dualities for the twisted torus, there are
two combinations of Killing vectors which lead to a closed isometry algebra, 
namely $\{k_1,k_2\}$ and $\{k_1,k_3\}$. Both choices result in a twisted T-fold:
\begin{itemize}

\item For a T-duality along Killing vectors $\{k_1,k_2\}$, the dual metric and $H$-flux 
are given by \eqref{dual_master_1} and \eqref{dual_master_2}, with the replacements
$h\leftrightarrow f$ and $R_1\to 1/R_1$.

\item For a T-duality along Killing vectors $\{k_1,k_3\}$, the expressions are similar
but now again with the additional changes $X^2 \leftrightarrow -X^3$ and $R_2 \leftrightarrow R_3$.

\end{itemize}


\subsubsection*{Three T-dualities}

The case of three collective T-dualities for the twisted torus is interesting since
here the isometry algebra is non-abelian.
However, due to the constraints \eqref{variations_45}, the $H$-flux has to vanish.
After applying the same formalism as above and performing the field
redefinitions
\eq{
  \tilde \chi_1 = \chi_1 \,, \hspace{40pt}
  \tilde \chi_2 = \chi_2 + f\op \chi_1 X^3\,, \hspace{40pt}
  \tilde \chi_3 = \chi_3 - f \op \chi_1 X^2 \,,  
}
we arrive at the following dual T-fold background
\begin{align}
  \nonumber
  &\check {\mathsf G} = 
  \frac{1}{R_1^2} \, d\tilde\chi_1 \wedge\star d\tilde\chi_1+
  \frac{1}{1+\bigl[\op \frac{f}{R_2R_3} \op  \tilde\chi_1 \bigr]^2}
  \left( \frac{1}{R_2^2} \, d\tilde\chi_2 \wedge\star d\tilde\chi_2
  +\frac{1}{R_3^2} \, d\tilde\chi_3 \wedge\star d\tilde\chi_3 \right)
  \,, \\[8pt]
  &\check {\mathsf H} = - \frac{f}{R_1^2R_2^2} \: 
  \frac{1-\bigl[\op \frac{f}{R_2R_3} \op  \tilde\chi_1 \bigr]^2}
  {\left(1+\bigl[\op \frac{f}{R_2R_3} \op  \tilde\chi_1 \bigr]^2\right)^2} \:d\tilde\chi^1 \wedge 
  d\tilde\chi_2\wedge d\tilde\chi_3\,.
\end{align}
Let us finally recall our discussion in section \ref{sec_recover}
about recovering the original model from the gauged action. We found that 
in the non-abelian case a change of basis characterized by a matrix $e_a{}^i$ has to be performed.
In the present case, this matrix takes the form
\eq{
  e_a{}^i = \left( \begin{array}{ccc}
  1 & 0 & - f X^2 \\
  0 & 1 & 0 \\
  0 & 0 & 1
  \end{array}\right) .
}


\end{appendix}


\clearpage
\bibliography{references}  
\bibliographystyle{utphys}


\end{document}